\documentclass[11pt,a4paper]{article}
\usepackage[utf8]{inputenc}
\usepackage[left=25mm,right=25mm,top=20mm,bottom=20mm]{geometry}

\usepackage[colorlinks,
linkcolor=blue,
anchorcolor=blue,
citecolor=blue,
urlcolor=blue]{hyperref}
\usepackage{graphicx}
\usepackage{lineno}
\usepackage{amsmath}
\usepackage{url}
\usepackage{booktabs}
\usepackage{array}
\usepackage{multirow}
\usepackage{tocbibind}
\usepackage{subfig}
\usepackage{caption}
\usepackage{xcolor}
\usepackage{xspace}
\usepackage{placeins}
\usepackage{verbatim}

\usepackage[utf8]{inputenc}
\usepackage[affil-it]{authblk}

\usepackage{soul,color,xcolor}
\setstcolor{red}

\begin{document}
    \title{
    Prediction of Energy Resolution in the JUNO experiment
    }
    \date{}
    
    \author[6,5]{Angel Abusleme}
\author[44]{Thomas Adam}
\author[47]{Kai Adamowicz}
\author[65]{Shakeel Ahmad}
\author[65]{Rizwan Ahmed}
\author[54]{Sebastiano Aiello}
\author[20]{Fengpeng An}
\author[22]{Qi An}
\author[54]{Giuseppe Andronico}
\author[66]{Nikolay Anfimov}
% ...
\affil[5]{Millennium Institute for SubAtomic Physics at the High-energy Frontier (SAPHIR), ANID, Chile}
\affil[6]{Pontificia Universidad Cat\'{o}lica de Chile, Santiago, Chile}
% ...
\affil[20]{Sun Yat-sen University, Guangzhou, China}
% ...
\affil[22]{University of Science and Technology of China, Hefei, China}
% ...
\affil[44]{IPHC, Universit\'{e} de Strasbourg, CNRS/IN2P3, F-67037 Strasbourg, France}
% ...
\affil[47]{III. Physikalisches Institut B, RWTH Aachen University, Aachen, Germany}
% ...
\affil[54]{INFN Catania and Dipartimento di Fisica e Astronomia dell'Universit\`{a} di Catania, Catania, Italy}
% ...
\affil[65]{Pakistan Institute of Nuclear Science and Technology, Islamabad, Pakistan}
\affil[66]{Joint Institute for Nuclear Research, Dubna, Russia}
% ...

    \maketitle
    Email: \href{mailto:Juno\_pub\_comm@juno.ihep.ac.cn}{Juno\_pub\_comm@juno.ihep.ac.cn}

    \begin{abstract}
    This paper presents an energy resolution study of the JUNO experiment, incorporating the latest knowledge acquired during the detector construction phase. The determination of neutrino mass ordering in JUNO requires an exceptional energy resolution better than 3\% at 1~MeV. To achieve this ambitious goal, significant efforts have been undertaken in the design and production of the key components of the JUNO detector.
    Various factors affecting the detection of inverse beta decay signals have an impact on the energy resolution, extending beyond the statistical fluctuations of the detected number of photons, such as the properties of the liquid scintillator, performance of photomultiplier tubes, and the energy reconstruction algorithm. To account for these effects, a full JUNO simulation and reconstruction approach is employed. This enables the modeling of all relevant effects and the evaluation of associated inputs to accurately estimate the energy resolution. The results of study reveal an energy resolution of 2.95\% at 1~MeV.
    Furthermore, this study assesses the contribution of major effects to the overall energy resolution budget. This analysis serves as a reference for interpreting future measurements of energy resolution during JUNO data collection. Moreover, it provides a guideline for comprehending the energy resolution characteristics of liquid scintillator-based detectors.
    \end{abstract}

\tableofcontents

\section{Introduction} 

The discovery of neutrino oscillations has established that at least two neutrino mass eigenstates are massive. 
This unambiguous conclusion has provided the first evidence of new physics beyond the Standard Model, which assumes neutrinos are massless. 
In the framework of three neutrino oscillations, six fundamental parameters are involved to describe the oscillation phenomena. 
Solar and reactor neutrino experiments have determined the mass-squared difference $\Delta m^{2}_{21}$ and mixing angle $\sin^2 \theta_{12}$.
Atmospheric and accelerator experiments have measured $|\Delta m^{2}_{32}|$ (or $|\Delta m^{2}_{31}|$) and $\sin^2 \theta_{23}$.
Reactor antineutrino experiments have provided the most precise determination of $\sin^2 \theta_{13}$ and have also achieved a comparable precision of $|\Delta m^{2}_{32}|$ with the measurements from atmospheric and accelerator experiments. 
The two independent mass-squared differences and the three mixing angles have been measured with a precision of a few percent~\cite{ParticleDataGroup:2022pth}. 
However, the sign of $\Delta m^{2}_{32}$ (referred to as neutrino mass ordering; NMO) and the value of the CP violating phase $\delta_{\textrm {CP}}$ remain partially unknown~\cite{T2K:2023smv, NOvA:2021nfi, Super-Kamiokande:2023ahc}; thus, several experiments~\cite{Hyper-Kamiokande:2022smq, DUNE:2021mtg, IceCube:2019dyb, KM3NeT:2021ozk} have been proposed to determine these two crucial parameters.  

The Jiangmen Underground Neutrino Observatory (JUNO)~\cite{JUNO:2015sjr,JUNO:2015zny,JUNO:2021vlw} in southern China is designed to determine the NMO by detecting reactor antineutrinos at baselines of 52.5~km from the Taishan and Yangjiang nuclear power plants.
The detector is equipped with a 20-kton spherical volume liquid scintillator (LS) located deep underground, providing an overburden of 650~m. 
JUNO aims to resolve the NMO by probing the interference effect of the two fast oscillations induced by $|\Delta m^{2}_{31}|$ and $|\Delta m^{2}_{32}|$ of reactor antineutrinos. 
Sufficient energy resolution is crucial to accurately measure the oscillation pattern in the antineutrino spectrum, and it directly determines the sensitivity to the NMO. In this context, the energy resolution refers to the sigma-to-mean ratio of the Gaussian function used to fit the energy distribution of positrons at a fixed energy.
The designed energy resolution of JUNO is 3\% at 1~MeV.
Taking this assumption as a benchmark, JUNO reported an NMO sensitivity in the design study at a confidence level of 3-4$\sigma$ after 6 years of data collection~\cite{JUNO:2015zny}. In addition, JUNO is expected to enable sub-percent-level precision measurements of the oscillation parameters $\sin^2 \theta_{12}$, $\Delta m^{2}_{21}$, and $\Delta m^{2}_{31}$~\cite{JUNO:2022mxj}. The experiment also serves as an observatory to detect neutrinos originating from both celestial and terrestrial sources~\cite{JUNO:2023zty,JUNO:2021tll,JUNO:2020hqc,JUNO:2022lpc}. Furthermore, the experiment is sensitive to the search for rare decays that go beyond the predictions of the Standard Model~\cite{JUNO:2022qgr}. Currently, JUNO is in the construction phase and is expected to start data collection in the near future.

Over the past few years several factors related to the NMO sensitivity in JUNO have undergone changes. As a result, the JUNO collaboration recently updated the NMO sensitivity using more realistic inputs~\cite{JUNO:2024jaw}. These updates include the incorporation of updated reactor thermal powers and new constraints on the reactor neutrino spectrum that we expect to obtain from the TAO~\cite{JUNO:2020ijm} satellite detector. Additionally, the accidental backgrounds have been revised based on radioassay results of the detector's raw materials and the estimation of radioactive impurities during the detector installation~\cite{JUNO:2021kxb}. The cosmogenic backgrounds have been re-evaluated due to changes in the overburden, and new backgrounds from global reactors and atmospheric neutrinos have been included. Furthermore, the detector energy resolution has been updated through full detector simulation and event reconstruction, which take into account a better understanding of the detector structure, as well as the performance of the LS and photomultiplier tubes (PMTs)~\cite{JUNO:2022hlz,Wang:2022tij}. This paper serves as a complementary report to the updated NMO sensitivity paper of JUNO and provides a detailed description of the latest understanding regarding the energy resolution prediction in the JUNO experiment, which is essential for NMO analysis.

Previous LS-based neutrino detectors, such as the Borexino~\cite{Borexino:2008gab} and KamLAND~\cite{KamLAND:2008dgz} experiments, have achieved energy resolution levels of 5\% and 6.5\% at 1~MeV, respectively. In these experiments, statistical fluctuations dominate the energy resolution, with reported photoelectron (PE) yields of 511~PE/MeV and 250~PE/MeV, respectively. The JUNO detector was designed with a focus on achieving high LS light yield, high transparency, and high photon collection efficiency to minimize the statistical fluctuation of the detected PE. 
The LS recipe has been optimized to maximize the light yield using one of the antineutrino detectors in the Daya Bay experiment~\cite{JUNO:2020bcl}. Additionally, 17,612 high-detection-efficiency 20-inch PMTs (LPMTs)~\cite{JUNO:2022hlz} and 25,600 3-inch PMTs (SPMTs)~\cite{Cao:2021wrq} have been deployed, resulting in a total photocathode coverage of 78\%. However, due to the unprecedented energy resolution requirement of JUNO, several other factors, which were usually negligible in previous experiments, significantly contribute to the energy resolution and must be carefully evaluated. These factors include the quenching effect in LS, Cherenkov radiation, electronics, and PMT responses. All of these effects have been modeled in the full detector simulation and will be thoroughly discussed in this paper, along with the impacts from the vertex and energy reconstruction algorithms.

The remainder of this article is organized as follows. We first provide an introduction to the JUNO detector in Sec.~\ref{sec:juno_detector}, followed by a summary of the main factors that impact the energy resolution throughout the processes of light production, propagation, and detection in Sec.~\ref{sec:origins_of_Ere}. The full detector simulation is then presented in Sec.~\ref{sec:full_detector_simulation}, focusing on the optical models of the LS and PMTs. Updates and improvements made to these models based on the latest knowledge and measurements are highlighted. The approaches used to determine key parameters in these models are also discussed. Additionally, the electronics simulation is introduced in Sec.~\ref{sec:elec}, which provides a comprehensive model of the detector response. Section~\ref{sec:energy_reconstruction} describes the calibration and event reconstruction procedures and reports the newly predicted energy resolution of JUNO, leading to a decomposition of the energy resolution to identify the major contributing factors in Sec.~\ref{decomposition} . Finally, a summary is given in Sec.~\ref{sec:summary}.

\section{The JUNO detector} \label{sec:juno_detector}

\begin{figure}[t]
    \centering
    \includegraphics[width=0.9\textwidth]{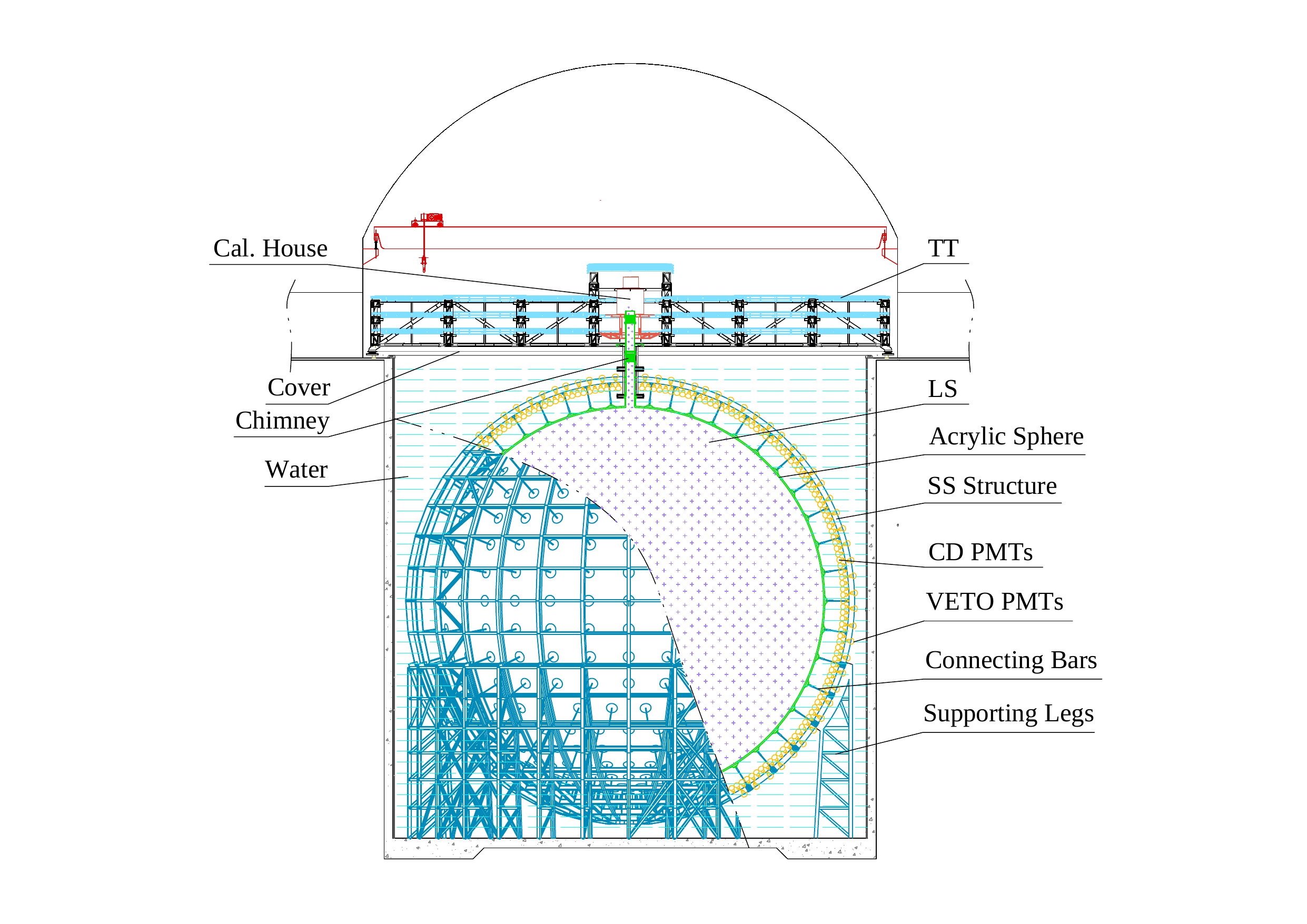}
    \caption{Schematic of the JUNO detector. TT: top tracker, LS: liquid scintillator, CD: central detector, SS: stainless steel.}
    \label{fig:juno_detector}
\end{figure}

A schematic of the JUNO detector is shown in Fig.~\ref{fig:juno_detector}. The detector consists of three main components: the central detector (CD), water pool (WP), and top tracker (TT).

The CD is located in the center of the WP and is composed of a 35.4~m inner diameter acrylic sphere with a wall thickness of 12.4~cm. The vessel is supported by 590 connecting bars and filled with 20 ktons of LS. The LS recipe consists of linear alkylbenzene (LAB) as the solvent, 2.5~g/L of 2,5-diphenyloxazole (PPO) as the fluor, and 3~mg/L of p-bis-(o-methylstyryl)-benzene (bis-MSB) as the wavelength shifter. Four purification plants have been constructed to ensure the desired radiopurity and target attenuation length. These plants allow for alumina filtration, distillation, water extraction, and gas stripping. The radiopurity of the LS will be monitored by the OSIRIS detector~\cite{JUNO:2021wzm} before and during the filling process.

Scintillation and Cherenkov light produced in the LS are collected by PMTs. PMT modules are the basic units used to assemble PMTs on the stainless steel (SS) supporting structure of the CD, accommodating multiple PMTs in each module. Among the LPMTs, 5,000 are dynode PMTs manufactured by Hamamatsu Photonics K.K. (HPK), while the remaining LPMTs are a new type of Micro-Channel Plate Photomultiplier Tubes (MCP-PMTs)~\cite{mcppmtintro} manufactured by Northern Night Vision Technology Co. (NNVT). Furthermore, a fraction of the NNVT MCP-PMTs, manufactured with improved technologies during photocathode fabrication, are labeled as NNVT HQE MCP-PMT. The arrangement of these LPMTs has been extensively studied, and their photocathode coverage has been determined to be 75\%. Each LPMT is equipped with a protection cover to prevent chain explosions in case of PMT breakage under the high water pressure. The protection consists of a front acrylic cover with a thickness of approximately 10~mm that follows the shape of the photocathode with a minimum 2~mm gap filled with water, working together with a back SS cover to provide full protection~\cite{He:2022qzj}. SPMTs are mounted in the gap between LPMTs, providing an additional $\sim$3\% photocathode coverage. The PMT modules made of SS provide optical isolation between the outer and inner WP. In addition, a comprehensive calibration system is implemented in the CD to control systematic errors in the energy scale and energy response of the detector~\cite{JUNO:2020xtj}.

The output signals from each LPMT are duplicated and amplified by two trans-impedance amplifiers (TIA) with different amplification factors. This configuration allows for a good signal-to-noise ratio for low-energy events, such as the inverse beta decay (IBD) process, with the high gain configuration. At the same time, it maintains a large dynamic range with the low gain configuration to detect high-energy events. The amplified analog signals from each TIA are then digitized using a custom flash analog-to-digital converter (FADC). The FADC provides a sampling rate of 1~GHz and resolution of 14 bits. The digitized waveform data are processed in the field programmable gate array (FPGA) to extract the charge and time information. The charge and time values of triggered events, along with the waveform data, are then sent to the data acquisition (DAQ) cluster. In the DAQ cluster, complex online event classification algorithms are employed to decide whether to save the full waveform or just the charge and time information onto the disk. This decision is made to optimize the data bandwidth and storage capacity. For IBD events, the waveform data are typically recorded to allow for more precise processing and analysis offline. More details of the LPMT electronics and trigger design are available in~\cite{Cerrone:2022dvp,Coppi:2023nlv,Triozzi:2023zct}. Regarding the SPMT electronics,  CatiROC chips are used to readout the signals, and only charge and time information are saved. More information is outlined in~\cite{JUNO:2020orn}.

The WP serves as a water Cherenkov detector and provides shielding from external radioactivity. The TT, located above the CD and WP, is a plastic scintillator detector used to tag cosmic muons~\cite{JUNO:2023cbw}. More detailed information on the JUNO detector can be found in~\cite{JUNO:2015sjr,JUNO:2021vlw}.

\section{Origin of the energy resolution} 
\label{sec:origins_of_Ere}

Reactor electron antineutrinos with energy exceeding 1.8~MeV can be detected within the LS target from interactions of the products of the IBD process: $\Bar{\nu}_e+p \to e^+ + n$. The positron ($e^+$), being much lighter than the neutron ($n$), carries away almost all the kinetic energy of the electron antineutrino. It quickly deposits its kinetic energy in the LS and annihilates into two 0.511\;MeV gammas, generating a prompt signal within an energy range of 1 to 12\;MeV. After scattering in the LS with an average lifetime of approximately $200~\mu$s, the neutron is captured by a hydrogen or carbon nucleus, consequently producing a delayed gamma signal of 2.22 or 4.95\;MeV, respectively. The coincidence signature of the prompt and delayed signals can effectively distinguish the IBD signals from the backgrounds. An accurate measurement of the positron energy and understanding of its energy resolution in the JUNO detector are crucial for NMO determination. The impact from the neutron recoils will not be discussed in this article, but it is considered in the NMO paper~\cite{JUNO:2024jaw}.

The energy deposition ($E_\mathrm{dep}$) of positrons primarily occurs through excitation and ionization processes. Because the concentrations of PPO and bis-MSB in the LS are relatively low, the deposited energy is mainly transferred to LAB molecules. Subsequently, the excited LAB molecules can transfer their energy to fluor molecules through complex molecular-scale processes~\cite{energy_transfer}. During the de-excitation of the fluor molecules, scintillation photons are emitted. However, the number of emitted scintillation photons is not proportional to the deposited energy due to the quenching phenomenon~\cite{Birks:1964zz}, where some excited molecules release energy without radiation emission. Various models exist to describe this quenching effect; the most common one is the semi-empirical formula proposed by Birks~\cite{Birks:1951boa}, which is used in this study:

\begin{equation} 
\label{eq:birks_formula}
    \frac{dN_\textrm{ScintOP}}{dx}=Y\frac{\frac{dE}{dx}}{1+kB\frac{dE}{dx}},
\end{equation}
where $N_\textrm{ScintOP}$ represents the number of optical photons produced by scintillation process, $Y$ corresponds to the scintillation light yield in units of photons per MeV without the presence of quenching, $dE/dx$ denotes the energy loss per unit path length (also known as the stopping power), and $kB$ is the Birks coefficient, which depends on the particle type. Given a fixed energy loss ($dE$) over a short path length of $dx$, $dN_\textrm{ScintOP}$ follows a Poisson distribution due to the random nature of the molecular-scale energy transfer process of scintillation, where the variance of the distribution is determined by $Y$. However, the total number of scintillation photons ($N_\textrm{ScintOP}$) produced by a particle no longer follows the Poisson distribution due to fluctuations in the energy loss described by a Landau distribution, the quenching effect modeled by Birks' Law, and the following random secondary particle generation processes:
\begin{itemize}
    \item Energetic $\delta$-electrons may be produced by ionization. They travel a significant distance away from the primary track and typically exhibit stronger quenching due to their higher $dE/dx$ compared to the primary positron.
    \item Gammas can be generated through Bremsstrahlung radiation and positron annihilation, and they interact through photoelectric effect, Compton scattering, and pair production. Then, the secondary electrons and positrons deposit their energies and produce scintillation photons through the aforementioned processes.
\end{itemize}

When the positron and its charged secondaries move through the LS at a velocity higher than the group velocity of light in the LS, Cherenkov photons are emitted. The Frank-Tamm formula~\cite{Frank:1937fk} is commonly used to calculate the number of Cherenkov photons produced per wavelength per unit path length  travelled by particles of charge $ze$:

\begin{equation}
\label{eq:frank-tamm}
    \frac{d^2N_\textrm{CherenOP}}{dx d\lambda}=\frac{2\pi\alpha z^2}{\lambda^2}\left(1-\frac{1}{\beta^2 n(\lambda)^2}\right).
\end{equation}
In this formula, $\alpha$ represents the fine structure constant, $\beta$ is the ratio of the particle's velocity to the speed of light in vacuum, and $n$ is the wavelength-dependent refractive index of the LS. According to Eq.~(\ref{eq:frank-tamm}), the number of Cherenkov photons is inversely proportional to the square of the wavelength ($\lambda$), resulting in a higher Cherenkov light yield at shorter wavelengths. However, there is very limited information on the refractive index at short wavelengths, which introduces significant uncertainties in estimating $N_\textrm{CherenOP}$. Although the Cherenkov process contributes additional light, it has a detrimental effect on the energy resolution. This is because the Cherenkov light yield depends on the path length of the charged particle in the LS, which can vary significantly due to the generation of secondary particles. This variation in path length can cause the Cherenkov photon distribution to deviate from Poisson statistics, leading to a degradation of the energy resolution.

Scintillation and Cherenkov photons produced in the LS may undergo various processes as they propagate through the detector, including absorption, scattering, reflection, and transmission at material boundaries. Absorption within the LS can be modeled as a competition among the LAB, PPO, and bis-MSB molecules. The absorption length of each component determines the probability of photon absorption. When photons are absorbed, they may be re-emitted again, and this re-emission probability is determined by the quantum yield (QY) of the fluors in the LS. Scattering within the LS is dominated by Rayleigh scattering rather than Mie scattering because LS is expected to have an extremely low concentration of large-sized impurities. The probability of Rayleigh scattering depends on the scattering length of the photons. The propagation of optical photons inside the LS can be well modeled with optical parameters such as absorption length, re-emission probability, Rayleigh scattering length, and refractive indices of the detector components. Only a fraction of these photons will be detected by the PMTs and converted into PE. The total PE number ($N_\textrm{PE}$) is influenced by factors such as the scintillation light yield, quenching effect, LS refractive index, light collection efficiency during transportation, and photon detection efficiency (PDE) of the PMTs. A simple energy reconstruction method is to divide $N_\textrm{PE}$, including contributions from both scintillation PE ($N_\textrm{ScintPE}$) and Cherenkov PE ($N_\textrm{CherenPE}$), by the energy scale, which is defined as the average PE number per MeV calibrated by 2.2~MeV gammas from neutron capture on hydrogen at the center of the CD. The average PE number produced in both scintillation and Cherenkov processes is not proportional to $E_\textrm{dep}$, leading to LS non-linearity (LSNL), denoted as
\begin{equation}
    f_\textrm{LSNL}(E_\textrm{dep})=\frac{E_\textrm{vis}(E_\textrm{dep})}{E_\textrm{dep}},
\end{equation}
where $E_\textrm{vis}$ is the visible energy defined as the expected reconstructed energy assuming a perfect energy resolution. Due to the aforementioned fluctuations in the energy loss and Cherenkov process, $N_\textrm{PE}$ does not follow a simple Poisson distribution but exhibits a larger standard deviation. To account for these fluctuations, a more general formula can be introduced as follows:
\begin{equation}
   \label{eq:MC_PE_resolution}
    \sigma_\textrm{PE}=\sqrt{\sigma_\textrm{ScintPE}^2 + \sigma_\textrm{CherenPE}^2 + 2\cdot cov[N_\textrm{ScintPE}, N_\textrm{CherenPE}]} > \sqrt{\langle N_\textrm{PE}\rangle},
\end{equation}
where $\sigma_\textrm{PE}$ is the standard deviation of the $N_\textrm{PE}$ distribution. Equation~(\ref{eq:MC_PE_resolution}) includes the correlation between the fluctuations in the scintillation and Cherenkov radiation which has been observed in the JUNO simulation and will be further discussed in the following section. The value of $N_\textrm{PE}$ also depends on the position of the IBD event. This position-dependent effect is referred to as the detector non-uniformity with respect to the amount of collected photons for the same energy release, which arises from geometrical factors. Consequently, the energy resolution in the JUNO detector is also position-dependent. 

The PMT and electronics responses can introduce additional fluctuations to the detected $\langle N_\textrm{PE}\rangle$. The PMT response includes various factors, such as the single photoelectron (SPE) resolution, which arises from the inherent fluctuations in the dynode or MCP charge amplification process. Additionally, effects such as dark count rate (DCR) and afterpulses can contribute to the fluctuations if they occur within the signal readout window and mix with the PE signal. The transit time and transit time spread (TTS) of the PMTs can impact the accuracy of vertex reconstruction and further influence the energy resolution due to the detector non-uniformity. On the electronics side, the presence of electronics noise can further smear the charge measurements. The digitization process can also contribute to the overall fluctuations. All of these factors and parameters have been well characterized during the PMT mass testing~\cite{JUNO:2022hlz} and electronics development phases~\cite{Bellato:2020lio}. Their impacts on the resolution depend also on the specific vertex and energy reconstruction algorithms employed.

Based on the waveform recorded in each readout channel, the charge and time information can be extracted by the waveform reconstruction algorithm. Then, the charge can be calibrated and converted to the PE number using PMT calibration algorithms. Eventually, the event position and energy can be obtained from the vertex and energy reconstruction algorithms using charge and time information of each readout channel as inputs. During reconstruction, additional fluctuations, PMT calibration, residual energy non-uniformity, and other factors are added to the reconstructed energy, where their amplitudes depend on the performance of the algorithms.

\begin{figure}
    \centering
    \includegraphics[scale=0.4]{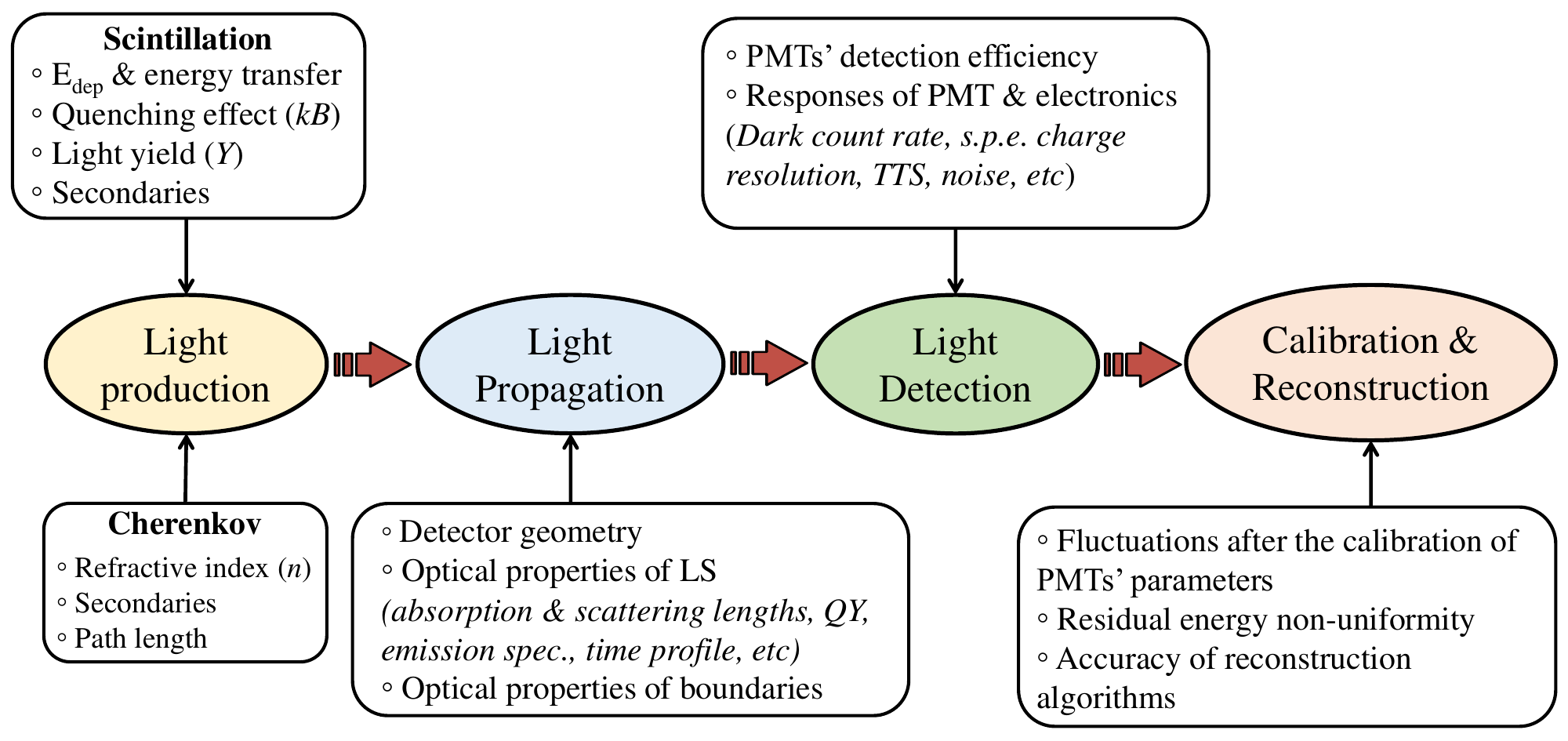}
    \caption{Summary of the key factors impacting the energy resolution throughout the processes of light production, propagation, detection, calibration, and reconstruction.} 
    \label{fig:resolution_flowchart}
\end{figure}

The main factors that may impact the energy resolution in the JUNO detector are summarized as follows (also illustrated in Fig.~\ref{fig:resolution_flowchart}):
     
\begin{enumerate}
     \item Positron energy loss fluctuation in LS and variation in its secondary particle generation in LS.
     \item Scintillation light yield and ionization quenching effect during scintillation photon production.
     \item Cherenkov photons emission.
     \item Light propagation and detection processes.
     \item PMT and electronics responses.
     \item Calibration scheme and energy reconstruction algorithm.
\end{enumerate}
All of the above items are carefully considered in the full JUNO detector simulation chain and algorithms of event reconstruction. Items 1-4, which include the energy deposition, scintillation (Sec.~\ref{sec:edep_and_quech}) and Cherenkov light production (Sec.~\ref{sec:cherenkov_radiation}), photon propagation (Sec.~\ref{sec:LS_properties_optical_model}), and detection (Sec.~\ref{sec:pmt_optical_model}), are taken into account in the simulation. Item 5, which involves the PMT and electronics responses (Sec.~\ref{sec:lpmt_elec_sim} and Sec.~\ref{sec:spmt_elec}), is modeled in a dedicated electronics simulation. Item 6 is included in the data processing of the calibration (Sec.~\ref{sec:calibration}) and reconstruction (Sec.~\ref{sec:rec_algo}).

\section{Full detector simulation} \label{sec:full_detector_simulation}
The JUNO detector simulation software was developed based on GEANT4~\cite{GEANT4:2002zbu, Allison:2006ve, Allison:2016lfl} version 10.04.p02. The simulation software, along with other offline data processing modules, is implemented within the Software for Non-collider Physics Experiment (SNiPER)~\cite{Zou:2015ioy, sniper} framework. Further details are outlined in~\cite{Lin:2022htc}. Aspects that potentially influence the energy resolution, which are modeled by the detector simulation, are discussed in the following sections. The aspects include the detector geometry, particle interactions in the detector media, light production, LS optical model describing the light propagation, and PMT optical model accounting for the reflection and angular-dependent PDE of the PMTs.

\subsection{Detector geometry modelling}
In detector simulation, mechanical design drawings, survey data, and the  best-known values for the optical properties of materials and  component dimensions are crucial to enable reliable simulation of light propagation, collection, and the detector non-uniformity response. The key modelled components include the following:
\begin{itemize}
    \item LS sphere with a radius of 17.7~m.
    \item Acrylic sphere with an inner radius of 17.7~m and thickness of 12.4~cm.
    \item Inner water buffer located between the acrylic sphere and PMT module with a thickness of 1.8~m.
    \item Detailed LPMT geometry~\cite{JUNO:2022hlz}, including photocathode, reflective aluminum film, inner electrode, and supporting structures, which are important for tracking photons inside the LPMTs. The 17,612 LPMTs are positioned in the inner water buffer facing the acrylic sphere, with a minimum distance of 1.42~m between acrylic and the center of the photocathode.  The acrylic and SS protection covers~\cite{He:2022qzj} are also modeled, located at the front and back of the LPMTs, respectively.
    \item 25,600 SPMTs~\cite{Cao:2021wrq} located within the gaps between LPMTs with a position distribution that guarantees a uniform SPMT density.
    \item 590 acrylic support nodes with detailed geometry and material definition, with corresponding SS support structure.
    \item PMT modules that optically isolate the inner and outer water pools.. 
    \item Calibration system~\cite{JUNO:2020xtj} with its anchors mounted on the acrylic sphere.
    \item Latticed SS shell used to support the LS acrylic sphere and mount the PMT modules and its own support structures.
    \item Outer water pool with a height and radius of 43.5~m.
    \item Chimney and calibration house on the top of the detector.
    \item TT consisting of 3 layers of plastic scintillator.
\end{itemize} 

\begin{figure}
    \centering
    \subfloat[]{
    \label{fig:other_rindex}
    \includegraphics[width=0.45\textwidth]{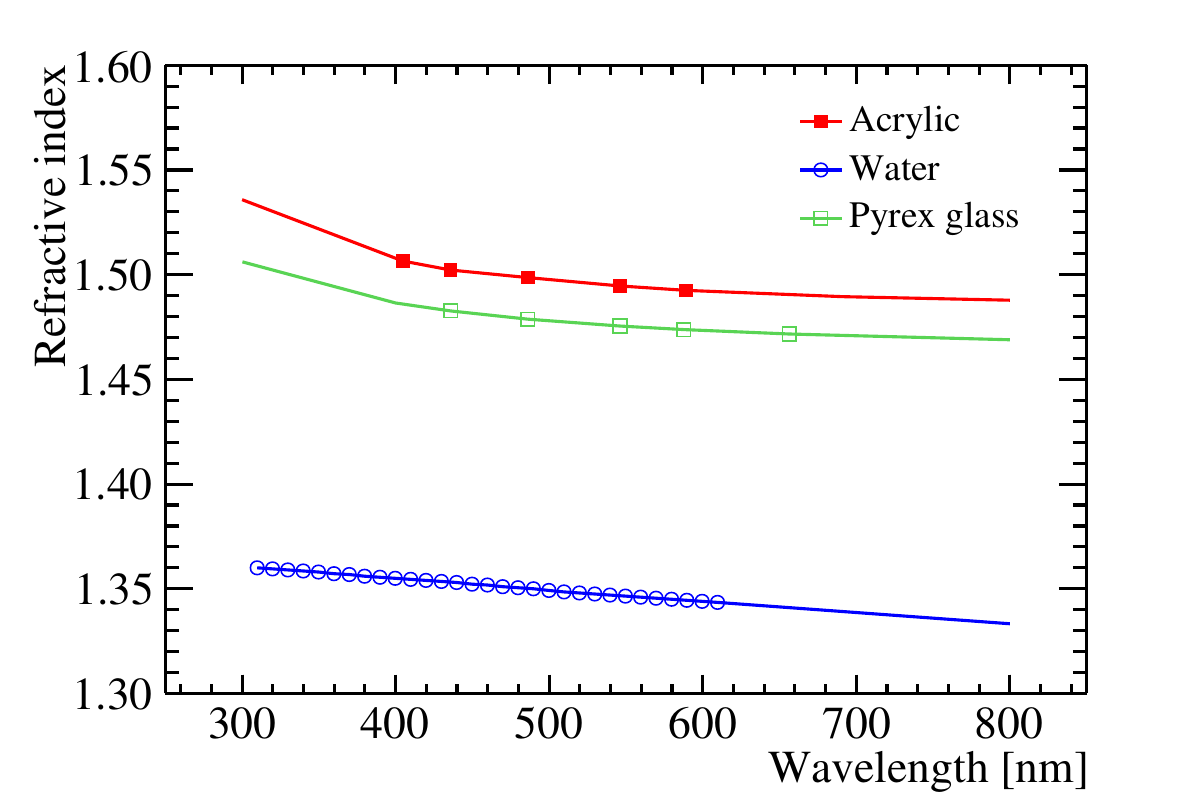}
    }
    \subfloat[]{
    \label{fig:other_att}
    \includegraphics[width=0.45\textwidth]{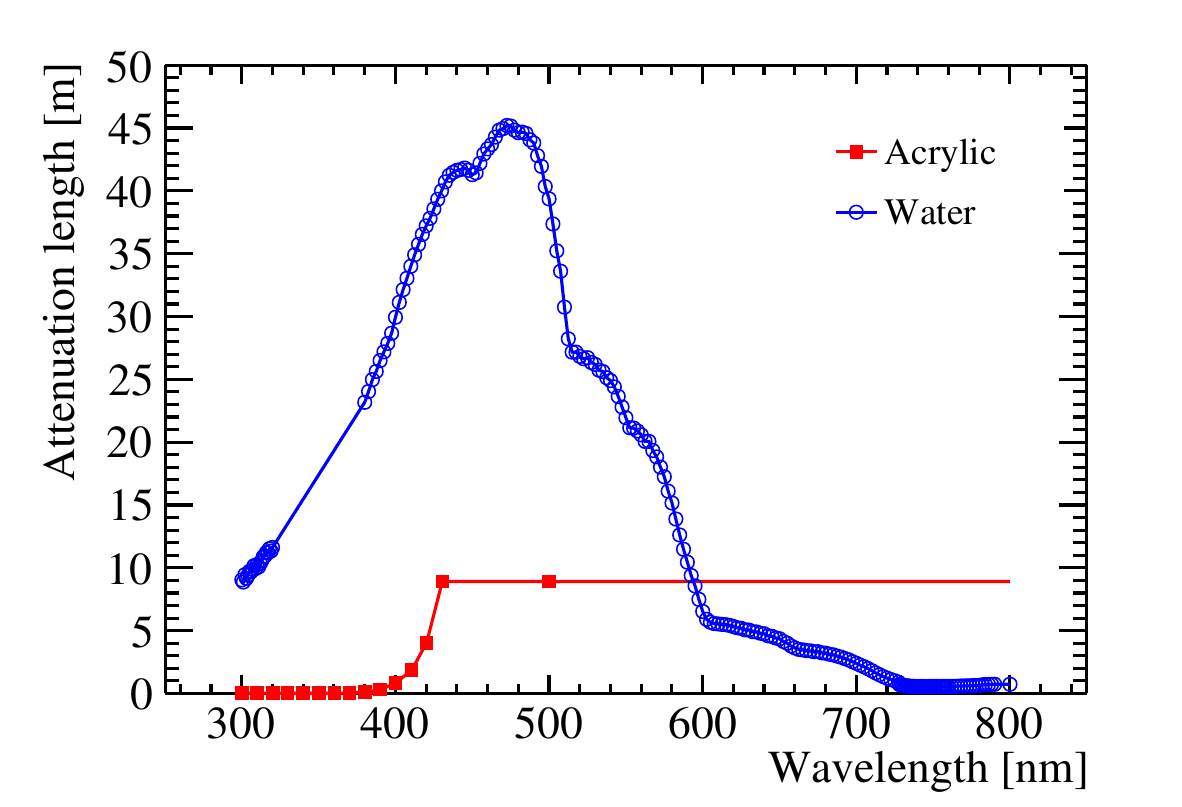}
    }
    \caption{(a) Refractive indices of acrylic (red), PMT glass (green), and water (blue) as a function of wavelength. (b) Attenuation lengths of acrylic (red) and water (blue) as a function of wavelength. Markers indicate measurements, while solid lines are interpolations or extrapolations.}
    \label{fig:other_rindex_att}
\end{figure}

The optical properties of each key component in the CD have been well defined in the simulation. In Figure~\ref{fig:other_rindex}, the refractive indices of the acrylic, water and PMTs' glass (Pyrex) are summarized as a function of wavelength. The refractive index of acrylic and glass is measured at 5 different wavelengths, indicated as markers. These measured values are then extrapolated to other wavelengths using the dispersion relation:
\begin{equation}
\label{eq:dispersion}
    n^2 - 1 = \frac{p_0 \lambda^2}{\lambda^2 - p_1},
\end{equation}
where $\lambda$ is the wavelength. $p_0$ and $p_1$ are two parameters obtained by fitting the 5 measured data points using Eq.~\ref{eq:dispersion}. Refractive index below 300~nm is not set in the simulation because a negligible number of photons from LS are expected to reach the acrylic or PMTs in this wavelength region, due to the strong absorption of LS. The refractive index of water was obtained from measurements in~\cite{Daimon:07}. The attenuation length of acrylic was obtained by analyzing the transmittance data published in~\cite{Yang:2021ecv, He:2024fdy}, shown as the red curve in Fig.~\ref{fig:other_att}. The attenuation length of water, shown as the blue curve in Fig.~\ref{fig:other_att}, is assumed to be 40~m at 430~nm, and its dependence on wavelength was taken from the Daya Bay collaboration. The absorption length is not set for the PMT glass because this effect is implicitly included in the PMTs' PDE. The reflectivity of the SS components is assumed to be 53.5\% without angular or spectral dependence, which is calculated using Fresnel’s equations with the refractive index obtained from~\cite{2017ApPhA123J}. 
This quantity will be measured in situ at JUNO and will be updated in the simulation in the future. The optical properties and modelling of the LS and PMTs will be discussed in more detail in Sec.~\ref{sec:LS_properties_optical_model} and \ref{sec:pmt_optical_model}, respectively. 

\subsection{Simulation of energy deposition and quenching effect} 
\label{sec:edep_and_quech}

\subsubsection{Energy deposition}
The low energy Livermore model~\cite{Apostolakis:1999bp} is chosen to simulate the electromagnetic processes of electrons, positrons, gammas, hadrons, and ions in the JUNO simulation. These processes include the photo-electric effect, Compton scattering, Rayleigh scattering, gamma conversion, Bremsstrahlung, and ionisation. The Livermore model is expected to possess improved simulation accuracy of energy deposition and secondary particle generation compared to the standard electromagnetic models because it directly uses shell cross-section data instead of parameterizations of these data. The Livermore model can handle particle interactions with energies down to 10~eV and is valid for elements with atomic number between 1 and 99. The physics list construction of \texttt{G4EmLivermorePhysics} provided by GEANT4 is directly used in the JUNO simulation. The production cuts are set to 1~mm for gammas and 0.1~mm for electrons and positrons, which are used to determine the energy thresholds of secondary particle generation during GEANT4 tracking. These cuts can significantly influence the determination of the Birks' coefficient. Beside the low-energy electromagnetic processes, a complete physics list has been constructed in simulation, including hadronic processes, ion physics, lepton and gamma-nuclear interactions, absorption, short-lived particles, radioactive decays, and optical processes. 

During simulation in GEANT4, the trajectory of a particle is tracked step by step. At each step, GEANT4 calculates the energy deposited by the particle, taking into account the fluctuations associated with the energy deposition process. The deposited energy in each step is then used to calculate the number of scintillation photons produced using Birks formula (Eq.~\ref{eq:birks_formula}) with the addition of Poisson fluctuations. The total number of scintillation photons produced in a physical event is obtained by summing over all the steps of both primary and secondary particles. 

\subsubsection{Determination of Birks' coefficient}
\label{sec:birks_parameter}
Constraining the Birks' coefficient ($kB$) through the energy non-linearity response of an LS-based detector, such as those measured by the Daya Bay and Borexino detectors~\cite{DayaBay:2019fje,Borexino:2013zhu}, can be challenging. This is because $kB$ is strongly correlated with the Cherenkov contribution, as discussed in the Daya Bay publication~\cite{DayaBay:2019fje}. However, it is possible to determine $kB$ by fitting the LS non-linearity data obtained from table-top measurements. These measurements typically involve electrons with energies below the Cherenkov threshold, typically around 0.2~MeV. Nevertheless, the $kB$ values reported in publications are not directly applicable to detector simulations because when extracting $kB$ using Eq.~\ref{eq:birks_formula}, only the $dE/dx$ of the primary particle is considered, usually calculated using tools like ESTAR~\cite{estar} or SRIM~\cite{srim}. This calculation does not account for the production of secondary particles. However, in the JUNO detector simulation, a significant number of secondary particles can be generated, carrying a fraction of the primary particle's energy. These secondary particles are independently tracked in GEANT4. To address this issue, a new fitting method has been explored to determine $kB$, which can be directly used in simulation. This fitting method takes into account the production and distribution of energy among secondary particles during simulation. 

\begin{figure}[htbp]
    \centering
    \includegraphics[width=0.6\textwidth]{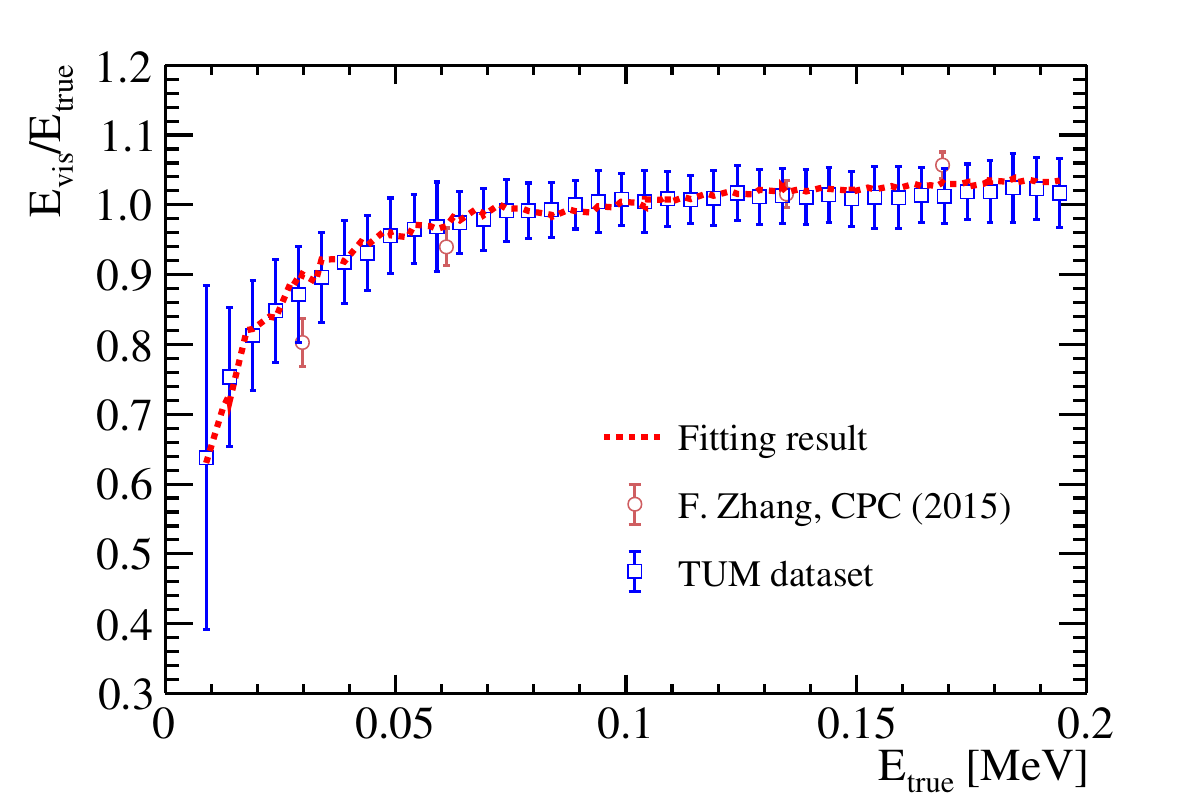}
    \caption{Combined fitting results on the electron quenching data below 0.2~MeV with production cuts of 1~mm for gammas and 0.1~mm for electrons and positrons. The data-set shown in brown color is from~\cite{Zhang_2015}. The data-set in blue is from the JUNO collaborators of TUM, where the systematic errors were estimated by measuring the quenching curves at four different incident angles of a $^{137}\textrm{Cs}$ gamma source with respect to the LS sample. The red curve is the fitting result using Eq.~(\ref{eq:chi2_func}).}
    \label{fig:electron_fit}
\end{figure}

The electron quenching effect of JUNO LS has been investigated by two groups: the Institute of High Energy Physics (IHEP) and the Technical University of Munich (TUM). Both groups used electrons generated from the Compton scattering of incident gammas emitted by radioactive sources. The gamma sources are placed outside a cylindrical LS container and can be rotated around it. The data-set marked with brown color was measured by the IHEP group~\cite{Zhang_2015}, in which a $^{22}\textrm{Na}$ source was used and only energies below 0.2~MeV were considered. The data set shown in blue was collected by the TUM group, where the gamma source of $^{137}\textrm{Cs}$ was rotated to change its position, and four different data-sets were obtained, corresponding to four different rotation angles. In Fig.~\ref{fig:electron_fit}, only one TUM data-set is shown, while the remaining data-sets are used to estimate the systematic errors.

At each data point in Fig.~\ref{fig:electron_fit}, electron interactions in the LS are simulated using GEANT4, with the initial electron energy $E_\textrm{true}$ derived from the Compton-scattered gamma. The production cuts in simulation are kept the same as those used in JUNO. It is important to note that the choice of production cuts can affect the energy deposition and consequently the value of $kB$. Therefore, it is crucial to keep the production cuts consistent throughout the study to ensure the reliability of the results. During the simulation, the energy deposition in each step is recorded. The visible energy, $E_\textrm{vis}$, for each event is calculated by summing up all the steps in the event, as indicated by Eq.~\ref{eq:birks_summation}, where $S$ is a fitting parameter and represents a normalization factor. A combined fit is then performed on the collected experimental data on electron quenching by minimizing the value of the $\chi^2$ function in Eq.~\ref{eq:chi2_func}:
\begin{align}
        E_\textrm{vis} &= \sum\limits_\textrm{step} \frac{S\cdot dE}{1+kB\left(\frac{dE}{dx}\right)} \label{eq:birks_summation}\\
        \chi^2 &= \sum_n\sum_i\left(\frac{\Bar{E}_\textrm{vis}/E_\textrm{true}-M_i^n}{\sigma_i^n}\right)^2, \label{eq:chi2_func}
\end{align}
where $n$ denotes the number of data-sets, $i$ refers to the number of data points in each data-set, ${E}_\textrm{vis}$ is the average visible energy of the simulation samples at the corresponding electron energy $E_\textrm{true}$, and $M$ indicates the measured ratio of $E_\textrm{vis}$ to $E_\textrm{true}$, with its corresponding error denoted by $\sigma$. Without loss of generality, we scale the $E_\textrm{vis}/E_\textrm{true}$ ratio to 1 at 0.1~MeV for each data-set. From the fitting, the Birks' parameter $kB$ is determined to be $(12.1\pm0.3)\times 10^{-3}\textrm{g}/\textrm{cm}^2/\textrm{MeV}$, with the uncertainty including both statistical and systematic errors.

\subsection{Cherenkov radiation} \label{sec:cherenkov_radiation}
\begin{figure}
    \centering
    \includegraphics[width=0.6\textwidth]{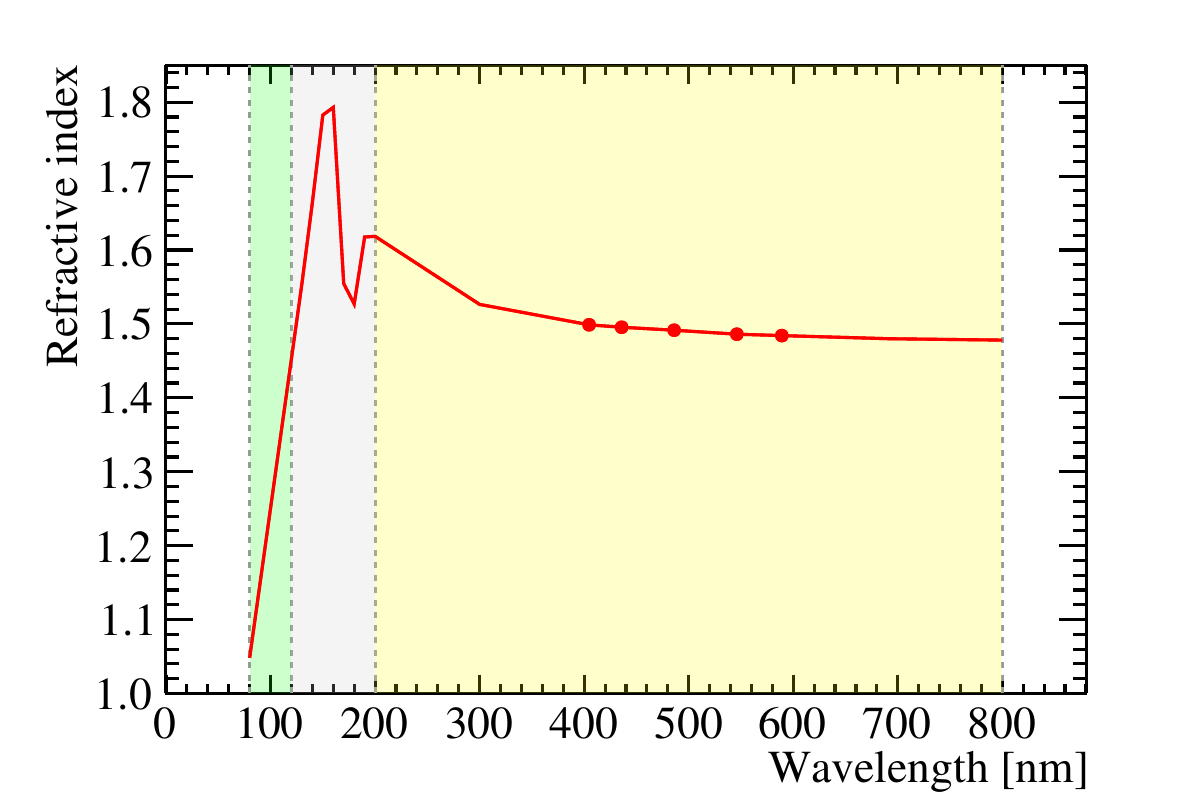}
    \caption{Refractive index of LS as a function of wavelength. The markers represent the data points measured by experiments using the V-prism refractometer~\cite{Zhou:2015gwa}. Then, the dispersion relation (Eq.~\ref{eq:dispersion}) is used to extend the refractive index down to 200~nm (yellow region). For wavelength range between 120~nm and 200~nm (gray region), the refractive index shape is taken from the KamLAND experiment. For wavelengths below 120~nm (green region), a linear extrapolation is utilized.}
    \label{fig:LS_rindex}
\end{figure}

The refractive index of LS is crucial for calculating the number of Cherenkov photons using Eq.~\ref{eq:frank-tamm}. In our simulation, we use the LS refractive index data shown in Fig.~\ref{fig:LS_rindex}, which were obtained with a precision better than 0.01\% at five different wavelengths (indicated as markers) using the V-prism refractometer~\cite{Zhou:2015gwa}. Then, we employ the dispersion relation given by Eq.~\ref{eq:dispersion} to extend the refractive index down to 200~nm. For the wavelength range between 120~nm and 200~nm, we adopt the refractive index shape from the KamLAND experiment~\cite{kamland_index} and scale it to match the refractive index of the JUNO LS at 200~nm. Finally, we perform a linear extrapolation to estimate the refractive index down to 80 nm, which has a value close to unity.

 Cherenkov photon production is handled by GEANT4 but with some modifications. This is necessary because GEANT4 assumes a refractive index that monotonically increases with photon energy, where the maximum refractive index corresponds to the maximum photon energy. However, this is not the case for the LS refractive index. To address this, we have enhanced the Cherenkov process in GEANT4 to handle more general forms of refractive index vs. photon energy curves. Initially, we use the photon energy range of the refractive index above the Cherenkov threshold based on the velocity information of incident particles. Subsequently, we calculate the number of Cherenkov photons for each energy range using Eq.~\ref{eq:frank-tamm}. Finally, we sample the energies of the emitted Cherenkov photons according to the LS refractive index curve.

It is essential to acknowledge that the refractive index and re-emission probability of the LS carry significant uncertainties, especially at shorter wavelengths, such as in the vacuum ultraviolet region ($<200$~nm). These uncertainties introduce a potential bias in predicting the Cherenkov light yield. To address this, we introduce a Cherenkov light yield factor, denoted as $f_C$, which is used to adjust the Cherenkov light yield in the simulation. The Cherenkov light yield factor is applied as follows:
\begin{equation}
\label{eq:Cherenkov_factor}
N_\textrm{CherenOP} = f_C \cdot N_\textrm{CherenOP}^\textrm{G4}.
\end{equation}
Here, $N_\textrm{CherenOP}^\textrm{G4}$ represents the calculated Cherenkov photon number obtained from GEANT4 using the LS refractive index as input. The factor $f_C$ is determined by constraining it with the LS energy non-linearity and energy scale measurements performed by the Daya Bay detectors. Further details regarding this constraint are discussed in Sec.~\ref{sec:abs_light_yield}.

\subsection{LS optical model and optical properties}
\label{sec:LS_properties_optical_model}
\begin{figure}[htbp]
    \centering
    \includegraphics[width=0.9\textwidth]{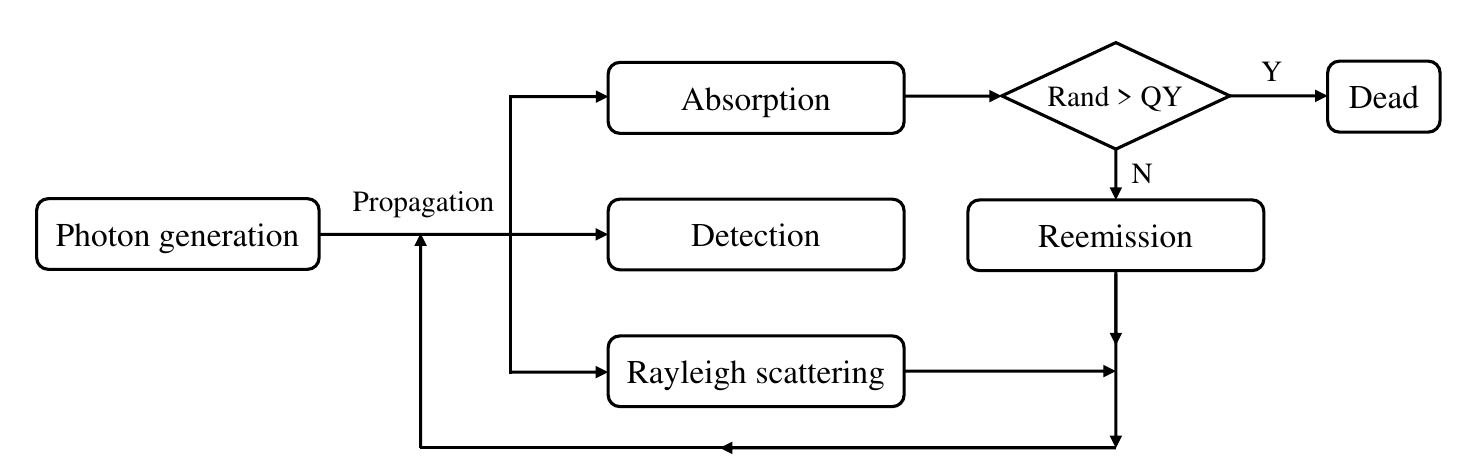}
    \caption{Schematic diagram of light propagation in the LS optical model.}
    \label{fig:LS_optical_model}
\end{figure}

Photon propagation in the LS is governed by the LS optical model, which takes into account the processes of emission, scattering, absorption, and re-emission. In this model, the three components of the LS, namely, LAB, PPO, and bis-MSB, are treated as a collective entity and share a set of equivalent optical parameters. These parameters include the photon emission spectrum, absorption and scattering lengths, and quantum yield after photon absorption. The LS optical model is illustrated in Fig.~\ref{fig:LS_optical_model}.
During the propagation of photons, they can either be absorbed or scattered by the LS, depending on the absorption and scattering lengths and their respective energy. If a photon is absorbed without undergoing the re-emission process, its trajectory is terminated. However, if re-emission occurs, a new photon is generated, and its energy is sampled from the LS emission spectrum. This newly generated photon continues its propagation within the LS.

The reflection and transmittance at the interfaces between the LS and acrylic, as well as between the acrylic and inner water buffer or between the water and PMT glass bulb, are accounted for by GEANT4 using the Fresnel formula and the predefined refractive indices of the LS, acrylic, water, and glass (Fig.~\ref{fig:other_rindex}). Additionally, photons have a probability of being absorbed by the acrylic or water, determined by their respective absorption lengths. Scattering within water uses Rayleigh scattering lengths calculated from the refractive index by GEANT4. Scattering within acrylic is currently neglected due to the lack of scattering length information. The boundary processes at the surfaces of the other detector components are also described by GEANT4, utilizing predefined optical properties such as reflectivity. After photon propagation, a fraction of the generated photons can impinge upon the PMT photocathodes, where they are further handled by the PMT optical model discussed in Sec.~\ref{sec:pmt_optical_model}.

\begin{figure}
    \centering
    \includegraphics[width=0.45\textwidth]{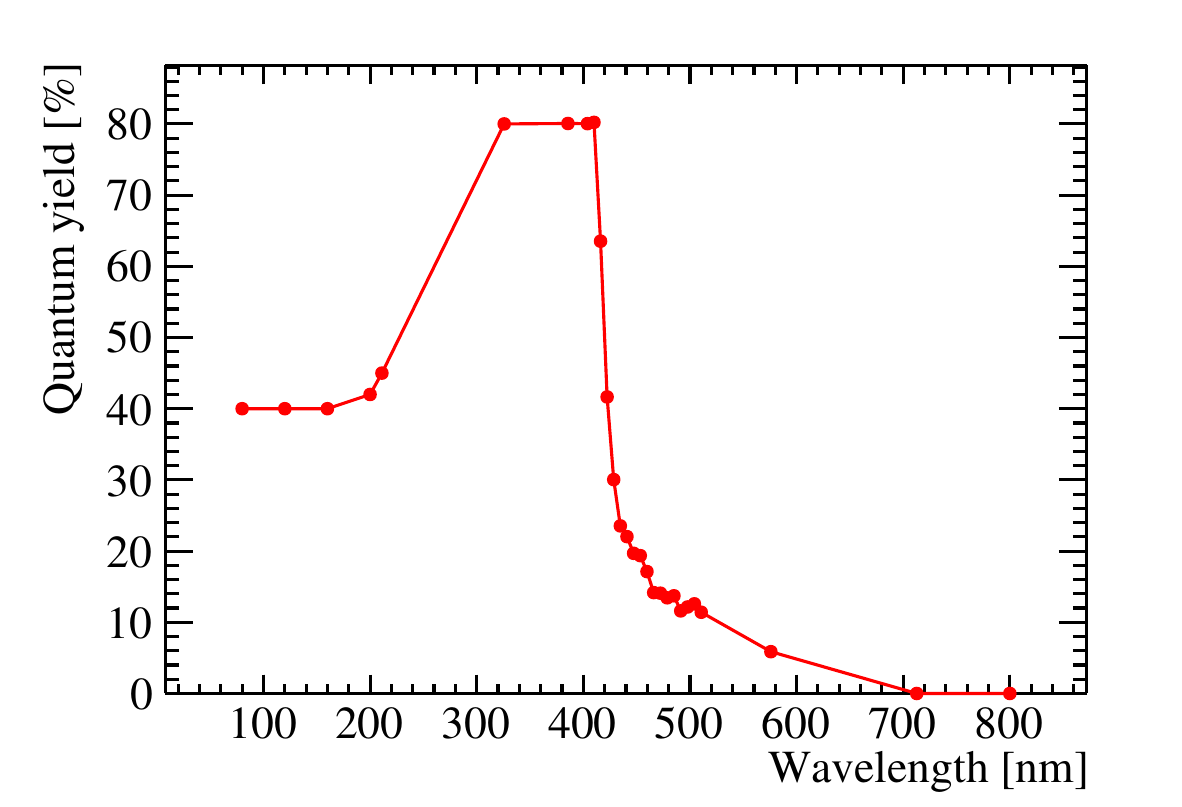}
    \includegraphics[width=0.45\textwidth]{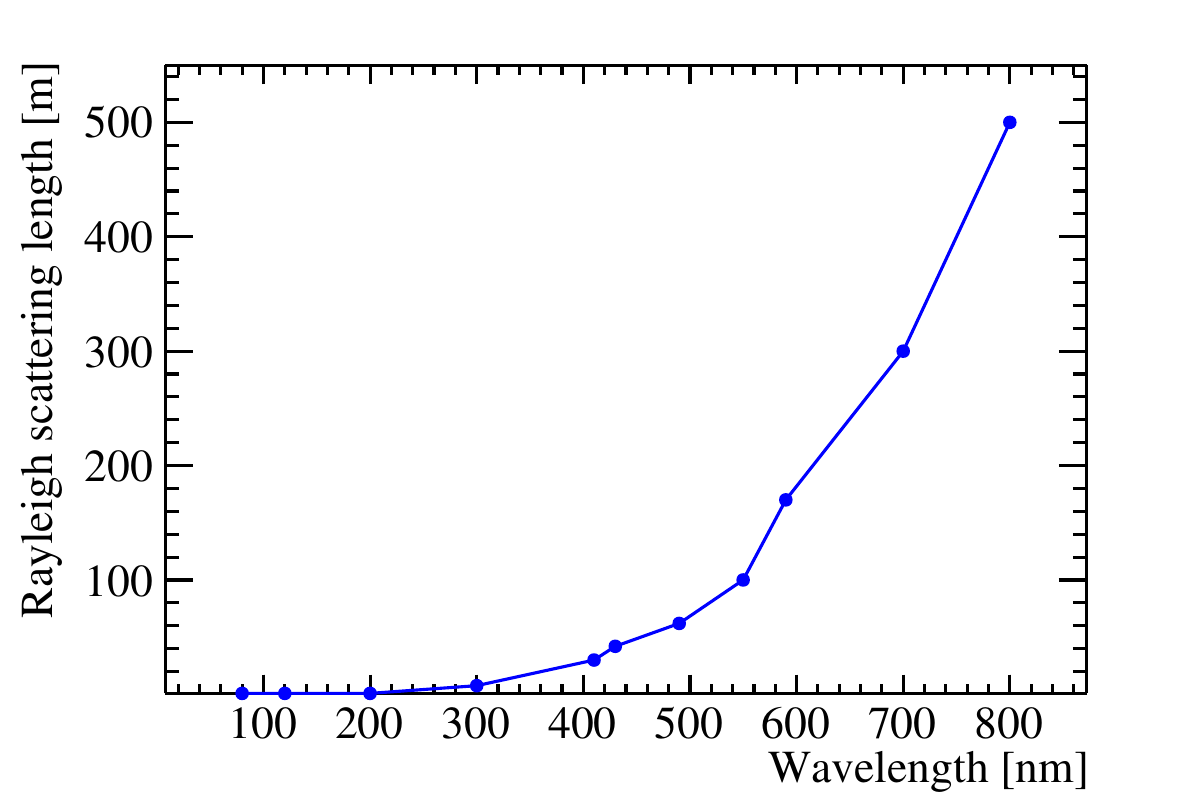}
    \caption{Quantum yield (left) and Rayleigh scattering length (right) as a function of wavelength.}
    \label{fig:QY_rayleigh}
\end{figure}

The LS optical properties employed in the LS optical model are obtained either from bench tests or inherited from the Daya Bay experiment. These properties are summarized as follows:
\begin{itemize}
    \item Emission spectrum: Considering the large size of the JUNO detector, the LS optical model employs the emission spectrum of bis-MSB. After undergoing several cycles of absorption and re-emission processes, the scintillation photons are expected to shift towards longer wavelengths and be primarily dominated by the emissions from bis-MSB, rather than PPO fluor. The bis-MSB emission spectrum is measured using a Fluorolog Tau-3 spectrometer, as shown in Fig.2 of~\cite{Zhang:2020mqz}. This spectrum is used to sample the energies of the scintillation photons during their production induced by ionization excitation and the re-emission process.

    \item Time profile: The time profiles employed in the simulation are obtained from dedicated measurements by exciting the LS with different charged particles, such as electrons, protons from neutron recoils, and alphas. The measured time profile is fitted using four exponential components, in which time constants of electrons/positrons/gammas are found to be 4.5~ns (70.7\%), 15.1~ns (20.5\%), 76.1~ns (6.0\%), and 397~ns (2.8\%)~\cite{LS_time_profile}. The time profile allows for the sampling of timing information for each scintillation photon. For the re-emission process, a single exponential component with a time constant of 1.5 ns, as measured in~\cite{Li_2011}, is employed.

    \item Rayleigh scattering length: The Rayleigh scattering length of the JUNO LS is obtained from measurements reported in~\cite{Zhou:2015gwa} and is shown in Fig.~\ref{fig:QY_rayleigh}, yielding a value of 27.0~m at 433~nm. This value is extrapolated to other wavelengths using the Einstein-Smoluchowski-Cabannes formula~\cite{TF9646001539}.

    \item Absorption length: The attenuation length ($L_{att}$) of the JUNO LS is assumed to have a target value of 20~m at 430~nm, which comprises both the absorption length ($L_{abs}$) and scattering length ($L_{sca}$). This can be expressed as $\frac{1}{L_{att}(\lambda)} = \frac{1}{L_{sca}(\lambda)}+\frac{1}{L_{abs}(\lambda)}$. Subtracting the measured scattering length from the attenuation length yields an absorption length of 77~m at 430~nm. The wavelength dependence of the absorption length is assumed to be the same as that of the Daya Bay LS~\cite{Goett:2011zz}.
    
    \item Quantum yield: The spectrum of the quantum yield was taken from the Daya Bay experiment, as shown in Fig.~\ref{fig:QY_rayleigh}, which has been fine-tuned to achieve agreement between the LS energy non-linearity in data and the simulation. The LS replacement experiment at Daya Bay~\cite{JUNO:2020bcl} indicates that the differences in PPO and bis-MSB concentrations between the Daya Bay LS (3~g/L PPO and 15~mg/L bis-MSB) and JUNO LS have a negligible impact on the quantum yield.
    \end{itemize}
    
The LS optical model, along with the optical properties, has been implemented in the detector simulation. The model is essential for a reliable simulation of photons in the detector, producing an accurate result on the light collection efficiency.

\subsection{Determination of the LS absolute light yield} 
\label{sec:abs_light_yield}
To predict the absolute light yield in LS, the two remaining unknown parameters, $Y$ in Eq.~\ref{eq:birks_formula} and $f_C$ in Eq.~\ref{eq:Cherenkov_factor}, can be constrained based on the LS energy non-linearity curve and energy scale measured by Daya Bay. This is done under the assumption that the scintillation light yield and energy non-linearity response are the same for both the Daya Bay and JUNO LS.

To ensure consistency between the JUNO and Daya Bay simulations, several modifications were made to the Daya Bay detector simulation as follows:
\begin{itemize}
    \item The same Livermore low-energy electromagnetic model is used, with the same production cuts.
    \item The same quenching effect model and $kB$ parameter discussed in Sec.~\ref{sec:birks_parameter} are employed.
    \item The LS refractive index and improved Cherenkov process are the same.
    \item The LS optical properties, such as the emission spectrum, quantum yield, scattering length, and time profiles, are assumed to be identical. The shape of the LS absorption spectrum is the same; however, the absolute absorption lengths differ. In Daya Bay, the absorption length is 27~m at 430~nm, as measured in the experiment.
    \item In Daya Bay, each PMT PDE is considered to be the measured quantum efficiency (QE), assuming a 100\% collection efficiency (CE). In JUNO, both QE and CE are taken into account, and their product, PDE at normal incidence, is constrained by the PMT mass testing data~\cite{JUNO:2022hlz}. The same PMT optical model (more details in Section~\ref{sec:pmt_optical_model}) is used to describe the PMT reflection and angular responses. However, different optical properties of the photocathode are applied for the 8-inch PMTs in Daya Bay and 20-inch PMTs in JUNO.
    \item The optical properties of other detector components in Daya Bay remain unchanged.
\end{itemize}

\begin{figure}
    \centering
    \subfloat[Comparison of the LS energy non-linearity between the Daya Bay simulation (red points) and Daya Bay calibration data (dashed line)~\cite{DayaBay:2019fje} with the tuned $f_C$. The shadow region indicates the error band of the data.]{
    \label{fig:reproduce_LSNL}
    \includegraphics[scale=0.4]{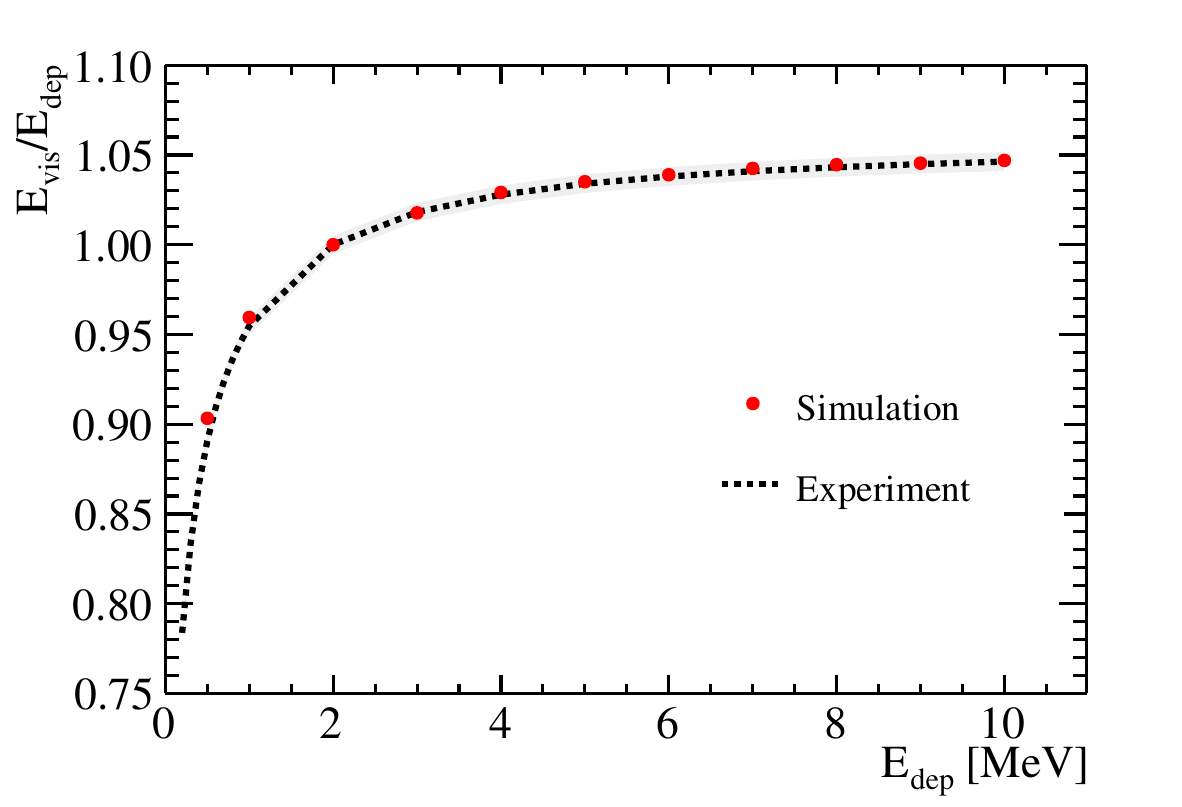}
    }
    \hfill
    \subfloat[$^{60}$Co energy spectrum at the detector center from the Daya Bay simulation before (blue shaded area) and after (red shaded area) $Y$ and $f_C$ scaling. The visible energy spectrum is fitted with a Crystal ball function combined with a Gaussian distribution (solid line), while the dashed line denotes the mean value of main peak of the simulated $^{60}$Co energy spectrum.]{
    \label{fig:reproduce_energy_scale}
    \includegraphics[scale=0.4]{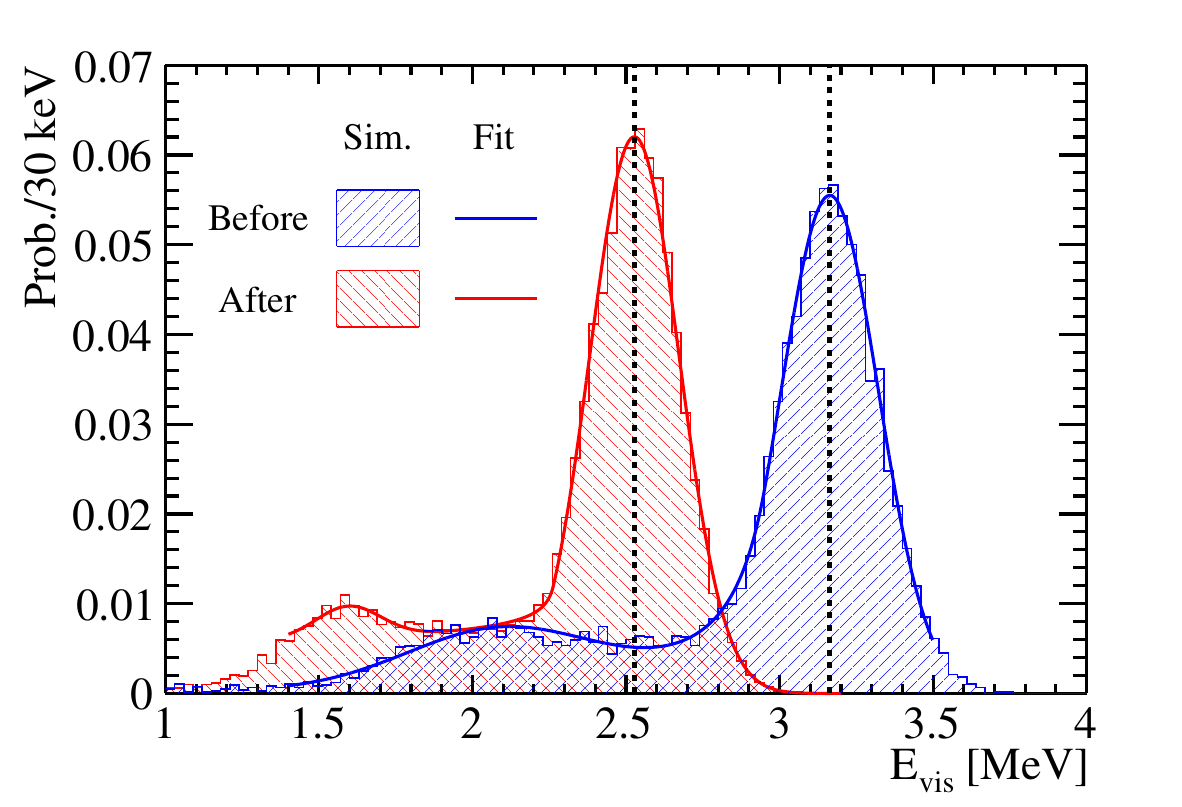}
    }
    \caption{Comparisons of the LS energy non-linearity and energy scale curves between the Daya Bay simulation and experimental data with the newly constrained parameters of $Y$ and $f_C$. }
    \label{fig:reproduce_es_LSNL}
\end{figure}

After making these modifications, the determination of the parameters $Y$ and $f_C$ in the Daya Bay simulation is carried out as follows:
\begin{enumerate}
    \item Fix the scintillation light yield $Y$ of the LS and tune $f_C$ in the Daya Bay simulation to match the gamma energy non-linearity curve in~\cite{DayaBay:2019fje}. By tuning $f_C$, the ratio of scintillation photons to Cherenkov photons, which determines the shape of the energy non-linearity curve, can be adjusted. Figure~\ref{fig:reproduce_LSNL} shows that good agreement can be achieved between the simulation and calibration data in Daya Bay.
    \item Simulate a $^{60}$Co radioactive source at the center of the detector in the Daya Bay simulation to obtain the visible energy. The energy scale in Daya Bay is defined as the average PE number per MeV using $^{60}$Co events at the detector center.
    \item Determine the value of $Y$ and $f_C$ by comparing the scaled visible energy to the expected value of $^{60}$Co, as shown in Fig.~\ref{fig:reproduce_energy_scale}. In this procedure, the ratio $f_C/Y$ is kept constant to ensure that the energy non-linearity response remains unchanged. Finally, $Y$ is determined to be 9846/MeV, while $f_C$ is 0.52.
\end{enumerate}

Table~\ref{tab:standard_parameters} summarizes a few parameters used in this work, allowing for the prediction of the detector energy resolution in JUNO.

\begin{table}[htbp]
    \centering
    \begin{tabular}{|c|c|c|c|c|}
    \hline
    \multirow{2}*{$Y$~[/MeV]} & \multirow{2}*{$kB~ [\textrm{g}/\textrm{cm}^2/\textrm{MeV}$]} & \multirow{2}*{$f_C$} & \multicolumn{2}{c|}{Production cut [mm]} \\
    \cline{4-5}
    &  &  & gamma & $e^+/e^-$ \\
    \hline
    9846 & 12.1$\times 10^{-3}$ & 0.52 & 1.0 & 0.1 \\
    \hline
    \end{tabular}
    \caption{Summary of a few parameters used in this work to predict the energy resolution in JUNO.}
    \label{tab:standard_parameters}
\end{table}

\subsection{PMT photon detection efficiency and optical model}
\label{sec:pmt_optical_model}
\subsubsection{LPMTs}
The PDE responses of LPMTs in JUNO are determined using the results from the LPMT mass testing setups~\cite{JUNO:2022hlz} and the developed PMT optical model~\cite{Wang:2022tij}. Two mass testing setups, the scanning station and container system, are used for LPMT acceptance tests and performance evaluation. The container system evaluates the characteristics of each LPMT, including PDE, DCR, gain, and features of SPE. A large area pulsed light source with a central wavelength of 420~nm is used to illuminate the entire photocathode of the LPMTs. The scanning station performs more detailed characterizations for LPMTs using 7 LED sources deployed along the longitude. The light beam from each LED is approximately perpendicular to the PMT surface and has a diameter of approximately 5~mm with a central wavelength of 420~nm. By rotating the PMTs, the whole photocathode can be scanned. However, even if only approximately 5\% of JUNO LPMTs are tested following this full scan procedure, the results are representative of the total JUNO LPMT production.

The number of PMT detected photons is assumed to follow a Poisson distribution. The average value measured with the container system is converted to the PDE defined by the scanning station system for all measured PMTs, because the light intensities in this system are calibrated by a reference PMT. The conversion coefficients are determined by comparing the PDEs measured by both the scanning station and container system for the three different types of LPMTs: HPK dynode-PMT, NNVT MCP-PMT, and NNVT HQE MCP-PMT. The PDE given by the scanning station is defined as the averaged value across the PMT area, and it is determined from PDEs measured by the 7 LEDs and their surface area weights. The area weights, calculated based on the respective positions on the photocathode of the 7 LEDs, are dependent on the LPMT types (HPK and NNVT) and are summarized in Table~\ref{tab:area_weight}.

\begin{table}
    \centering
    \caption{Surface area weights of 7 LEDs for NNVT and HPK PMTs.}
    \begin{tabular}{|c|c|c|c|c|c|c|c|}
    \hline
      & LED1 & LED2 & LED3 & LED4 & LED5 & LED6 & LED7 \\
    \hline
      NNVT & 4.8\% & 9.0\% & 12.6\% & 17.2\% & 20.0\% & 18.0\% & 18.4\% \\
    \hline
      HPK  & 4.5\% & 8.8\% & 13.5\% & 17.1\% & 20.5\% & 18.6\% & 17.0\% \\
    \hline
    \end{tabular}
    \label{tab:area_weight}
\end{table}

\begin{figure}[ht]
    \centering
    \subfloat[]{
        \label{fig:PMTs_CE}
        \includegraphics[width=0.45\textwidth]{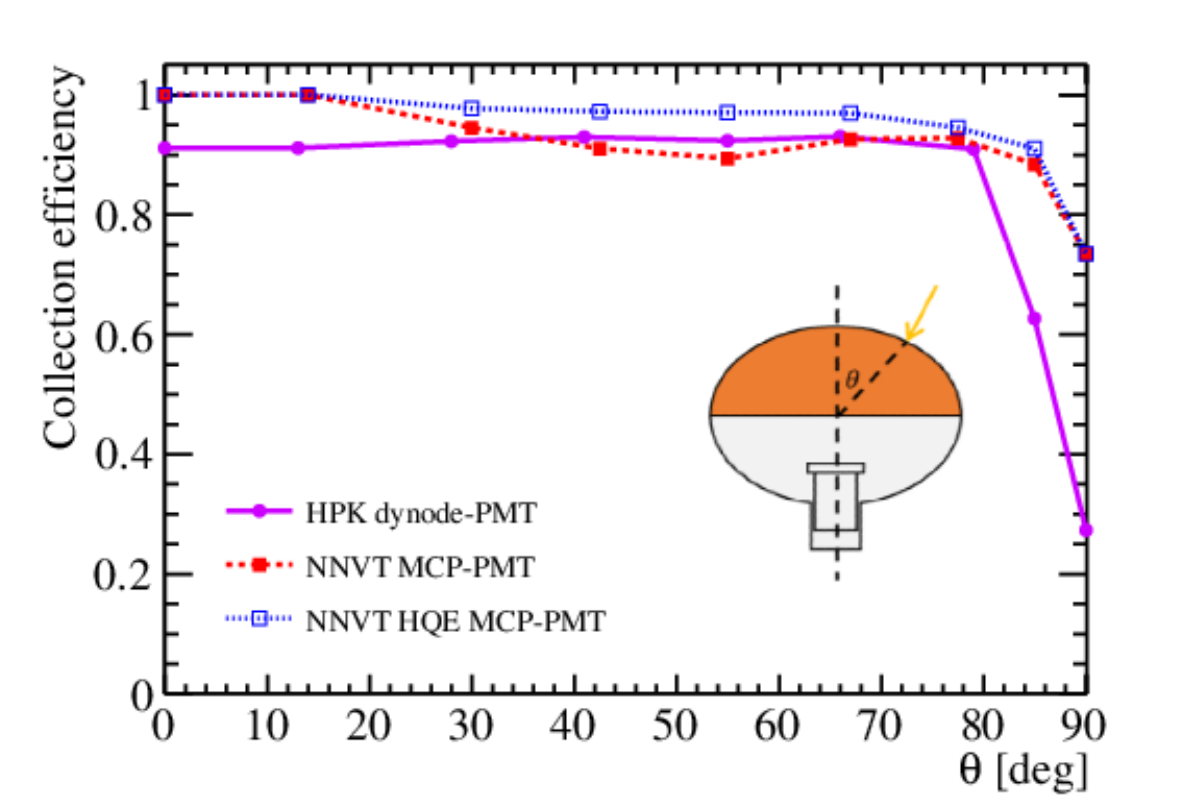}
    }
    \subfloat[]{
        \label{fig:PMTs_QE}
        \includegraphics[width=0.45\textwidth]{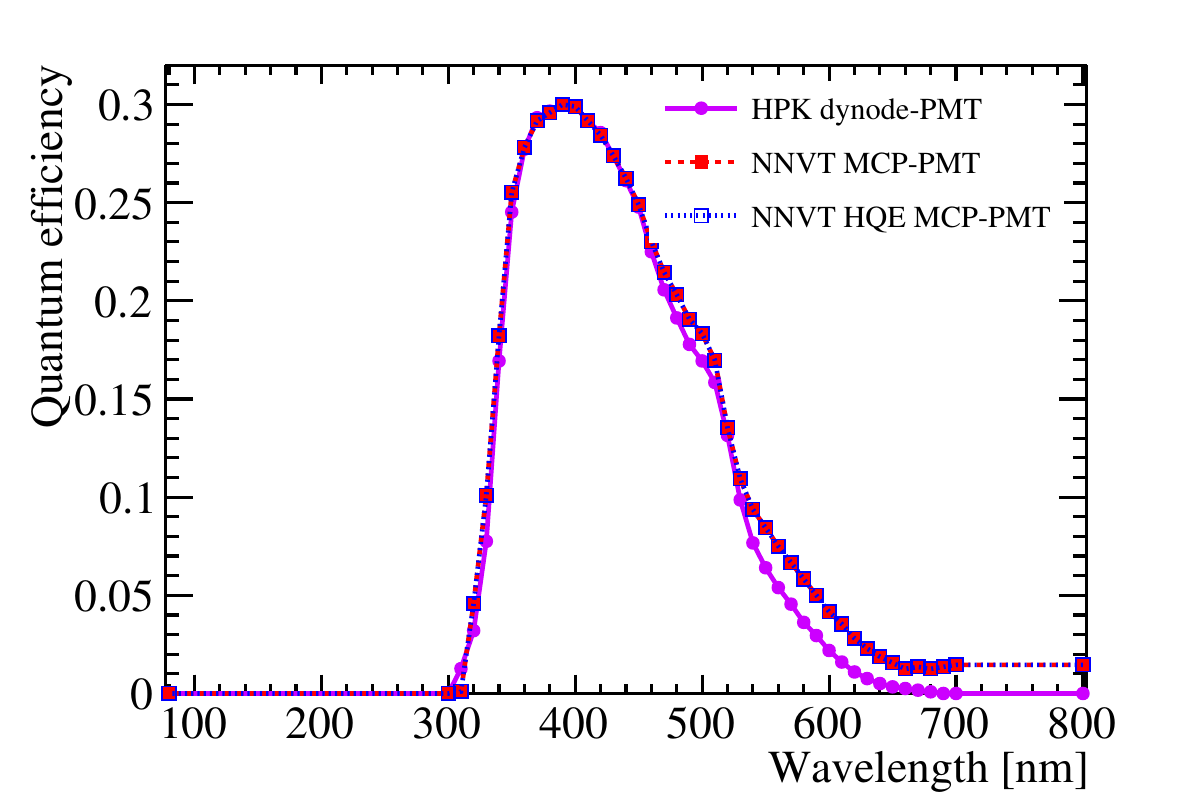}
    }
    \caption{(a) Angular dependence of collection efficiency for NNVT MCP-PMT (red), NNVT HQE MCP-PMT (blue), and HPK dynode-PMT (violet). (b) Wavelength dependence of quantum efficiency for NNVT MCP-PMT (red), NNVT HQE MCP-PMT (blue), and HPK dynode-PMT (violet).}
    \label{fig:PMT_eff}
\end{figure}

In the detector simulation, the position dependence of the PDE along the latitude of the photocathode is modeled using the CE curves. The PDE is considered as a product of the QE and CE. Figure~\ref{fig:PMTs_CE} shows the CE as a function of the polar angle in the PMT's local coordinates for the NNVT MCP-PMT (red), NNVT HQE MCP-PMT (blue), and HPK dynode-PMT (violet). The CE curves are obtained by averaging the PDEs measured by the 7 LEDs along the zenith angles in the scanning station. The maximum CE values are set to 100\% for the NNVT MCP-PMT and 93\% for the HPK dynode-PMT, indicated by the electrostatic simulation results from NNVT and HPK private communications, respectively.

To compute the QE of a given LPMT, the following equation is used:
\begin{equation}
    \textrm{QE}=\textrm{PDE}/\sum_{i=1}^{7}(\textrm{CE}_i\times w_i),
\end{equation}
where $i$ represents the index of the LED in the scanning station, and $w_i$ denotes the surface area weight of the $i$-th LED, as summarized in Table~\ref{tab:area_weight}. This calculation is performed for each LPMT in the CD to ensure that the product of QE and CE is consistent with the measured PDE, as discussed in~\cite{JUNO:2022hlz}. The QE spectral responses are also implemented in the simulation based on laboratory measurements, as shown in Fig.~\ref{fig:PMTs_QE}. The NNVT MCP-PMT (red) and NNVT HQE MCP-PMT (blue) have identical QE spectral responses, while the HPK dynode-PMTs (violet) have a different response at higher wavelength ($\lambda>500~\mathrm{nm}$). It is assumed that LPMTs of the same type share the same CE curve and QE spectral response in the simulation.

The PMT optical model~\cite{Wang:2022tij} takes into account the angle of incidence (AOI) dependence of the PDE, as well as the reflections on the photocathode and optical processes inside the PMTs. In this model, the PMT window is treated as a multi-layer optical stack, from outside to inside, consisting of water, PMT glass, anti-reflective coating, photocathode, and vacuum layers. The anti-reflective coating and photocathode are considered coherent layers due to their comparable thicknesses with the light wavelength. The model incorporates the light interference effect and the multiple reflections between adjacent boundaries using the transfer matrix method.

The refractive index $n$, extinction coefficient $k$, and thickness $d$ of the anti-reflective coating and photocathode are determined in the wavelength range of 390~nm to 500~nm by analyzing the reflectance data of NNVT MCP-PMT, NNVT HQE MCP-PMT, and HPK PMT immersed in the LAB liquid. The optical properties of the other components inside the PMTs, such as dynode (MCP), supporting structure, and aluminum film, are constrained by the QE data. These parameters are presented in~\cite{Wang:2022tij} and are directly used in this work.
With these inputs, the PMT optical model calculates both the reflectance and absorbance for a given photon and AOI, assuming uniformity for the anti-reflective coating and photocathode. The reflected photons are then propagated by GEANT4. The absorbance at a specific AOI is converted to the QE using the QE calculated above and absorbance information at normal incidence. This optical model was integrated into the detector simulation and used in this work. It has also been applied in the Daya Bay simulation, using the optical parameters obtained from the reflectance and QE data of an 8-inch Daya Bay PMT.

\subsubsection{SPMTs}
In the detector simulation, a simple PMT optical model is used for SPMTs. It assumes that photons hitting the photocathode are 100\% absorbed and converted to free PE by applying the PMTs' QE. The QE at 420~nm is implemented for each SPMT, using the value obtained from the characterization of SPMTs and published in~\cite{Cao:2021wrq}. The QE dependence on wavelength is considered according to the vendor's datasheet~\cite{sPMT_QE}.

\subsection{PE yield}
In the updated detector simulation, the predicted PE yield of the JUNO detector is 1665~PE/MeV, calibrated using neutron capture events on hydrogen at the detector center. The PE yield is larger compared to the previously reported value of 1345~PE/MeV~\cite{JUNO:2020xtj}. This enhancement can be attributed to three factors:
\begin{itemize}
    \item Improved PMT PDE: In previous studies, the LPMT's PDE was assumed to be 27\%. However, the LPMT mass testing has shown that the actual average PDE in CD is 30.1\%~\cite{JUNO:2022hlz}. This higher efficiency accounts for an $\sim$11\% increase in PE yield.
    \item More realistic PMT optical model: Previous simulations used a simplified PMT optical model that neglected the angular dependence of PDE in water and PMT photocathode reflection. By incorporating a more realistic optical model~\cite{Wang:2022tij}, we observed an additional $\sim$8\% increase in PE yield.
    \item Detector geometry updates: The detector geometry is updated based on the final mechanical design, with reflections on several detector components taken into account, leading to an approximate $\sim$3\% increase in PE yield.
\end{itemize}

Furthermore, the Cherenkov to total PE ratio is found to be 1.1\% for 1~MeV positrons, and the detector simulation indicates a PE yield of 1785~PE/MeV for uniformly distributed neutron capture events on hydrogen.

\section{Electronics simulation} 
\label{sec:elec}
The LPMT and SPMT electronics simulations are important for predicting the energy resolution of the detector, as they contribute to the overall electronic noise and affect the signal-to-noise ratio of the recorded signals. These simulations are integrated into the SNiPER framework to provide a comprehensive modeling of the detector response.

\subsection{LPMT electronics simulation} \label{sec:lpmt_elec_sim}
\subsubsection{LPMT contribution} \label{sec:lpmt_response}
In the LPMT electronics simulation, LPMT operation parameters are obtained from the PMT mass testing systems and are directly assigned to each LPMT using its serial number. These characteristics include the dark count rate (DCR), gain, and SPE charge resolution.

To simulate the LPMT dark noise, SPE pulses are uniformly distributed in a readout time window of $\Delta T=1\mu s$. The number of dark noise pulses is sampled from a Poisson distribution with mean value $\textrm{DCR}\times\Delta T$.

\begin{figure}[htbp]
    \centering
    \subfloat[Charge spectrum of MCP-PMT]{
    \label{fig:MCP_spe_spectrum}
    \includegraphics[width=0.45\textwidth]{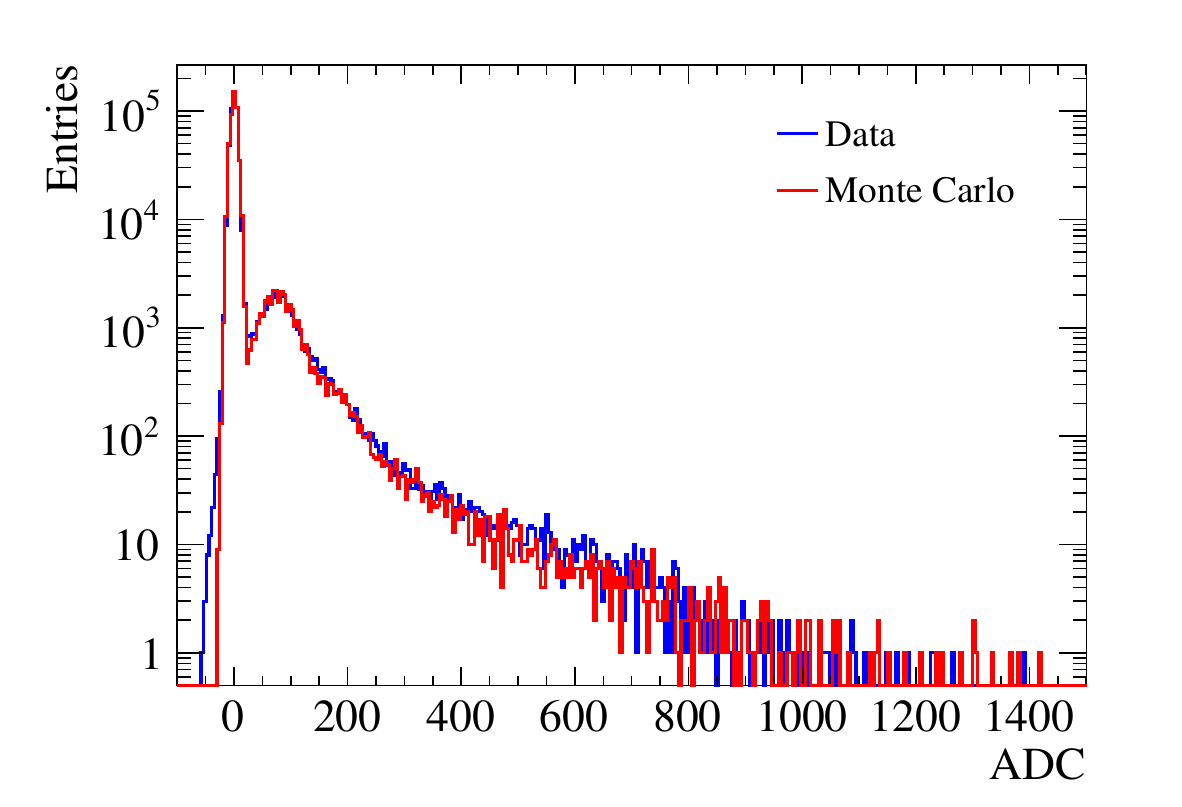}}
    \subfloat[Fraction of large signals]{
    \label{fig:large_signal_ratio}
    \includegraphics[width=0.45\textwidth]{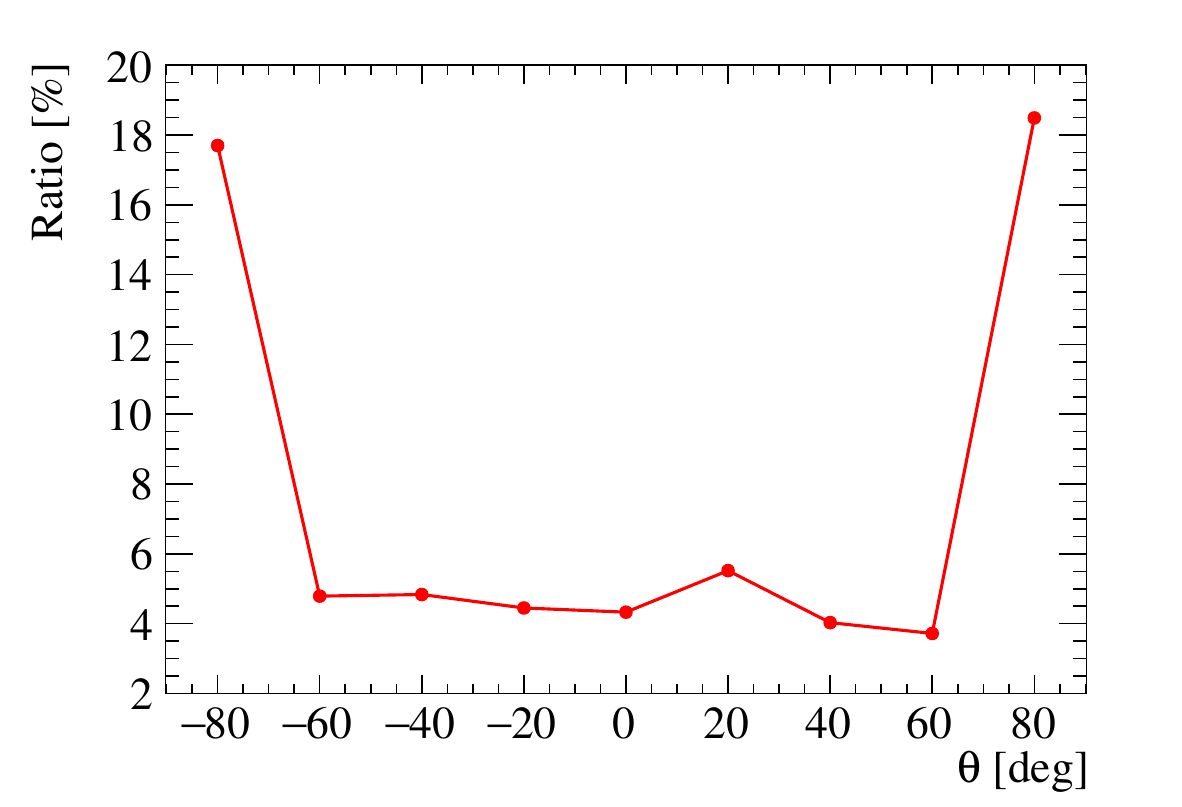}}
    \caption{(a) The charge spectrum of NNVT MCP-PMT, measured with a pulsed light source, exhibits a long tail, which is described by an exponential distribution in the Monte Carlo simulation. Good agreement between data and Monte Carlo is achieved. (b) The ratio of the large signals depends on the positions on the photocathode, and is extracted from experimental data.}
    \label{fig:MCP_charge_spectrum}
\end{figure}

When converting an SPE from the detector simulation into a pulse in the electronics simulation, the amplitude of the pulse is modeled by a combined distribution of a Gaussian and exponential function. The SPE charge resolution of each PMT, obtained from the PMT mass testing system~\cite{JUNO:2022hlz}, is directly used as the sigma of the Gaussian component in the simulation. The exponential component exhibits distinct behaviors between HPK LPMTs and NNVT LPMTs. For the former, the contribution from the exponential component is found to be 1\% with an average amplitude of 1.1 PE, resulting in an SPE relative variance of 0.4.  For the latter, the exponential component describes the large signal observed in the measured charge spectrum of MCP-PMT (blue line in Fig.~\ref{fig:MCP_spe_spectrum}) with an average amplitude of 2.2 PE, resulting in an SPE relative variance of 0.7. The ratio of large signals in the SPE spectrum of MCP-PMT, as a function of the zenith angle on the photocathode, has been characterized by laboratory measurements, as shown in Fig.~\ref{fig:large_signal_ratio}. Good agreement between Monte Carlo simulation and experimental data is achieved, as illustrated in Fig.~\ref{fig:MCP_spe_spectrum}.

The transit time and TTS are also implemented for both HPK and NNVT LPMTs. As these quantities are only measured for approximately $25\%$ of LPMTs~\cite{JUNO:2022hlz}, a random sampling based on the distributions is used to obtain transit time and TTS values for other LPMTs in the CD. The transit time and TTS values are position-dependent on the photocathode, and this effect is also considered in the electronics simulation based on dedicated measurements made during the PMT mass testing.

The after-pulses of LPMTs are modeled based on the results published in~\cite{Zhao:2022gks}. As the earliest after-pulse component occurs at approximately 1~$\mu$s after the primary signals, the overlap with the signals is expected to be small and thus has a negligible impact on the energy resolution. The non-linearity effect of LPMTs is also taken into account in the electronics simulation. However, it primarily affects high-energy events rather than IBD ones.

\subsubsection{Electronics contribution}
In the LPMT electronics simulation, the SPE average waveform is obtained using the PMT testing data for both NNVT and HPK LPMTs. This waveform serves as a template in the electronics simulation to model the SPE response. The amplitude of the waveform is determined by the gain and SPE charge resolution discussed earlier.

To model the overshoot effect, the same model as that used in the Daya Bay experiment~\cite{Jetter:2012xp} is adopted, but with parameters tuned specifically for the JUNO experiment. The overshoot model is parameterized using an exponential function plus a Gaussian function. To capture the onset of the overshoot, the exponential function is multiplied by a Fermi function. The maximum amplitude of the overshoot is set to be 1\% of the primary pulse~\cite{Luo:2016ddb}.

White noise is applied to the raw waveform by sampling a Gaussian function with a mean of 10\% of the SPE amplitude. This value is based on the requirements of the JUNO experiment. The digitization process is also modeled in the simulation, taking into account the effects of the front-end analog-to-digital converter (FADC) resolution, high-gain and low-gain amplifications of the SPE signal, and baseline offset. The simulation also considers the non-linearity effects of the electronics. However, it is anticipated that these effects will have a negligible impact on the energy resolution for positrons coming from IBD events.

Overall, the LPMT electronics simulation in the CD takes into account various aspects of the LPMT response, including the SPE waveform, overshoot, white noise, digitization, and non-linearity effects. These simulations aim to accurately model the electrical responses of the LPMTs and readout electronics and their impact on the energy resolution of the detector.

\subsection{SPMT electronics simulation} \label{sec:spmt_elec}
The SPMT electronics simulation aims to accurately model the response of SPMTs and the readout electronics. For each SPMT, a DCR value is randomly sampled and assigned based on the DCR distribution obtained from the SPMT mass production measurements~\cite{Cao:2021wrq}. The mean values of the TTS (1.6~ns) and SPE charge resolution (33\%) are shared by all SPMTs.

The key features of the CatiROC ASIC are implemented in the simulation based on the measurements in Ref.~\cite{JUNO:2020orn}, including the two types of dead times: 70~ns trigger dead time and 9.3~$\mu$s digitization dead time. In cases where a hit is not triggered due to dead time, a fraction of its charge can still be calculated. This fraction is called the charge acceptance and is parameterized based on the time difference with the previous hit. In addition, the charge in picocoulombs is converted into an ADC unit, with two conversion coefficients that responsible for high gain (less than 10~PE) and low gain (more than 10~PE).

\section{Reconstruction} \label{sec:energy_reconstruction}
\subsection{Reconstruction and calibration of PMT waveforms} \label{sec:calibration}
As described in Sec.~\ref{sec:elec}, the PMT together with the electronics system will convert the photons to waveforms for each LPMT and to the pairs of charge and time for each SPMT.
For each LPMT waveform, the charge and hit time of every pulse are reconstructed using the deconvolution algorithm~\cite{Huang:2017abb}, which exhibits smaller charge non-linearity compared to other algorithms.
Meanwhile, the hit time of the first pulse, referred to as the first hit time, is of particular importance. A dedicated algorithm with a linear fit of the rising edge of the first pulse is developed to achieve a more accurate first hit time and reduce its charge dependence.
Given that the charge and hit time information of PMTs are the inputs to the vertex and energy reconstruction in Sec.~\ref{sec:rec_algo}, the performance of the waveform reconstruction will also contribute to the energy resolution.
Moreover, the CD has approximately O(10$^4$) PMTs, and their characteristics may differ. Thus, the reconstructed charge and hit time of each PMT must be calibrated to account for the different PMT parameters, such as gain, TTS, relative PDE, and DCR.
A comprehensive calibration strategy was developed in Ref.~\cite{JUNO:2020xtj} to extract these parameters for all the PMTs and continuously monitor their time dependence.

\subsection{Event reconstruction methodology and results}
\label{sec:rec_algo}

The aim of the event reconstruction is to derive the energy and vertex of the event from the charge and time information of photon hits captured by PMTs.
However, for large-volume LS detectors such as JUNO, various complex optical processes occur during the photon propagation. 
It is usually relatively difficult to build a comprehensive optical model to precisely describe the photon hits of PMTs. 

Given that a non-uniform detector response is one of the main contributors to the energy resolution for large LS detectors, precise vertex reconstruction is needed to correct for the energy response non-uniformity.
Several algorithms have been developed for vertex reconstruction in JUNO~\cite{Liu:2018fpq, Li:2021oos}. These algorithms utilize the time information of the first photon hit of PMTs, together with the residual time probability distribution function (pdf), which is mainly determined by the LS timing profile and PMTs' TTS. 
Meanwhile, an optical-model-independent method~\cite{Wu:2018zwk} was developed to reconstruct the event energy in JUNO and was optimized in Ref.~\cite{Huang:2021baf} to further improve the energy uniformity. The basic principle is to obtain the expected charge for PMTs using calibration data from the automatic calibration unit (ACU) and cable loop system (CLS), which is then used to build a likelihood function given the observed charge of all PMTs. 

A calibration data-driven simultaneous vertex and energy reconstruction method has been developed for JUNO~\cite{Huang:2022zum}, based on the vertex and energy reconstruction methods described above.
The observables considered are the charge and time information of the PMTs.
Calibration data with known vertices and energy are used to construct the expected charge and time response for each PMT. 
Given the observed and expected charge and time information of PMTs, a maximum likelihood method is developed  to reconstruct the event vertex $\mathbf{r}$=$\vec{r}(r,\theta,\phi)$ and visible energy $E_\textrm{}$ simultaneously, using the likelihood function in Eq.~\ref{eq:EqQTMLE}:

\begin{eqnarray}
\label{eq:EqQTMLE}
    \begin{aligned}
    \mathcal{L}(\{q_{i}\}; \{t^{j}\}|\mathbf{r},E_{\textrm{}}, t_0) =& 
        \prod_{i} \left(\sum_{k=1}^{\infty} P_{Q}(q_{i}|k) \times P(k, \mu_i) \right) \\ 
&\times \prod_{j} \frac{\sum^{K}_{k=1} P_{T}(t^j_{res}|r, d_j, \mu^{l}_j, \mu^{d}_j, k)\times P(k, \mu^{l}_j+\mu^{d}_j)} {\sum^{K}_{k=1} P(k, \mu^{l}_j+\mu^{d}_j)},       \\
\end{aligned}
\end{eqnarray}
\begin{eqnarray}
\begin{aligned}
    &\mu_i(\mathbf{r},E_{\textrm{}}) = E_{\textrm{}}  \times \hat{\mu_i}^L(r,\theta,\theta_{\text{PMT},i}) + \mu_i^D \\
    &t^{j}_{res} = t^{j} - t^{j}_{tof}(d_j) - t_0 \\
    \end{aligned}
\end{eqnarray}

The first product on the right side of Eq.~\ref{eq:EqQTMLE} corresponds to the charge-based likelihood function. The index $i$ runs over all PMTs.
The term in parentheses simply describes the probability of observing charge $q_i$ on PMT $i$ when the expected number of PEs is $\mu_i$, 
which strongly depends on both the vertex and energy of the positron. 
Given that photons emitted from the same particle have strong temporal correlation and usually arrive on PMTs within a few hundred ns, while PMT dark noise occurs randomly in time, a signal window of 420~ns is set to reduce the PE contamination from dark noise.
$\mu_i^D$ represents the residual PE contribution from dark noise within the signal window.
As one of the most crucial ingredients of the reconstruction, $\hat{\mu_i}^L(r,\theta,\theta_{\textrm{PMT},i})$ represents the expected number of LS PEs per unit of visible energy, originating from particles within the signal window. This is obtained using the simulated calibration data, in which the calibration source is deployed at various positions in the CD.
Here, $r$ and $\theta$ are the components of the particle vertex $\mathbf{r}$, while $\theta_{\textrm{PMT},i}$ is the angle between $\mathbf{r}$ and the PMT position vector $\mathbf{r}_{\textrm{PMT},i}$.
$P(k, \mu_i)$ is the Poisson probability for detecting $k$ PEs,
$P_{Q}(q_{i}|k)$ is the probability of observing charge $q_i$ on PMT $i$ given the charge pdf of $k$ PE $P_{Q}(q|k)$, which can be constructed by convolving the SPE charge spectrum with $P_{Q}(q|k-1)$.
Note that for PMTs that do not pass the firing threshold of $q_i > 0.1~\textrm{PE}$, this term simplifies to $P(0, \mu_i)$+$P_{Q}(q_i<0.1PE|k=1)$*$P(1, \mu_i)$. Moreover, the index $k$ ends when $P_Q(q_i|k)<10^{-8}$ is met to simplify the calculation.

The second product on the right side of Eq.~\ref{eq:EqQTMLE} corresponds to the time-based likelihood function. 
The index $j$ only runs over the fired PMTs satisfying $-100~\textrm{ns} <t^j_{res} < 500~\textrm{ns}$ and $0.1~\textrm{PE} < q_{j} < K$. A cutoff value of $K=20$ is set for the detected nPE $k$ to simplify the calculation.
The residual hit time $t_{res}$ of PMTs is obtained by subtracting the time of flight $t_{tof}$ and reference time $t_0$ from the first hit time $t$ of PMTs.
The distance between the vertex and PMT is denoted as $d$.
$\mu^l$ and $\mu^d$ represent the expected number of PEs originating from particle or dark noise within the full electronic readout window, respectively. Another crucial ingredient of the reconstruction is the residual time PDF $P_{T}(t^j_{res}|r, d_j, \mu^{l}_j, \mu^{d}_j, k)$, which is obtained using the same calibration data used for $\hat{\mu_i}^L$. The fine-grained parameterization of $P_{T}(t^j_{res})$ takes into account its dependence on the vertex radius as well as the distance between the vertex and PMT. Meanwhile, the impact from dark noise is also included via an analytical approach. More details of the construction of $P_{T}(t^j_{res})$ can be found in Ref.~\cite{Huang:2022zum}. $P(k, \mu^{l}_j+\mu^{d}_j)$ acts as a weight for different $k$ values. 

Compared to previous reconstruction studies, a few important updates should be mentioned.
First, two crucial ingredients have been made more realistic: the residual time pdfs are derived from calibration data instead of MC simulation, and all PMT electronics effects are considered for the construction of the expected nPE map $\mu_i$. Second, the fine-grained parameterization of the residual time pdf as well as the calibration of the time of flight as a function of the photon propagation distance makes the pdf more accurate. Third, the charge and time information of PMTs are combined together to improve the vertex resolution, especially near the acrylic sphere edge. Finally, the vertex and energy are reconstructed simultaneously, which naturally handles the strong correlation between these two quantities.

To evaluate the performance of the energy reconstruction in JUNO, a few sets of positron samples with different kinetic energies $E_k$ = (0, 0.5, 1, 2, 3, 4, 5, 6, 8, 11) MeV were produced by MC simulation, as summarized in Tab.~\ref{tab:PhySampleInfo}. For each data-set, the positrons are uniformly distributed in the CD, and the total statistics per set is 500k.
In addition, the $^{68}$Ge calibration data listed in Tab.~\ref{tab:PhySampleInfo} were also produced to obtain the expected nPE map and time pdfs of PMTs. 
The positions and type of calibration source have been slightly optimized based on Ref.~\cite{Huang:2021baf} to improve the energy uniformity.
A realistic detector geometry with all the latest knowledge on the properties of LS and PMTs from previous sections were implemented in the simulation. 
For each set of positrons with fixed kinetic energy $E^i_k$, the simultaneous vertex and energy reconstruction using Eq.~\ref{eq:EqQTMLE} was applied. 
The distribution of the reconstructed visible energy $E_\textrm{rec}$ was fitted with a Gaussian function ($E^i_\textrm{vis}, \sigma^i$), and the results are summarized in Tab.~\ref{tab:E_rec}. The corresponding energy resolution is defined as the ratio of  $\sigma^i$/$E^i_\textrm{vis}$, and the energy non-linearity is calculated by $E^i_\textrm{vis}$/${E^i_\textrm{dep}}$, where ${E^i_\textrm{dep}}$ = ${E^i_{k}} + 1.022~\textrm{MeV}$.

\begin{table}[!ht]
\centering
\caption{List of MC simulation samples.}
    \begin{tabular}{lccc}
\toprule
       Type     & \textbf{Energy} & \textbf{Statistics} & \textbf{Position}\\
\midrule
     $e^+$ & $E_k$=(0, 0.5, 1, 2, 3, 4, 5, 8, 11)~MeV& 500k/set & uniform in CD \\
     $^{68}$Ge  & $0.511\cdot2~$MeV &  20k/point & ACU+CLS (293 points)  \\
\bottomrule
\end{tabular}
\label{tab:PhySampleInfo}
\end{table}

\begin{figure}
    \centering
    \includegraphics[width=0.45\textwidth]{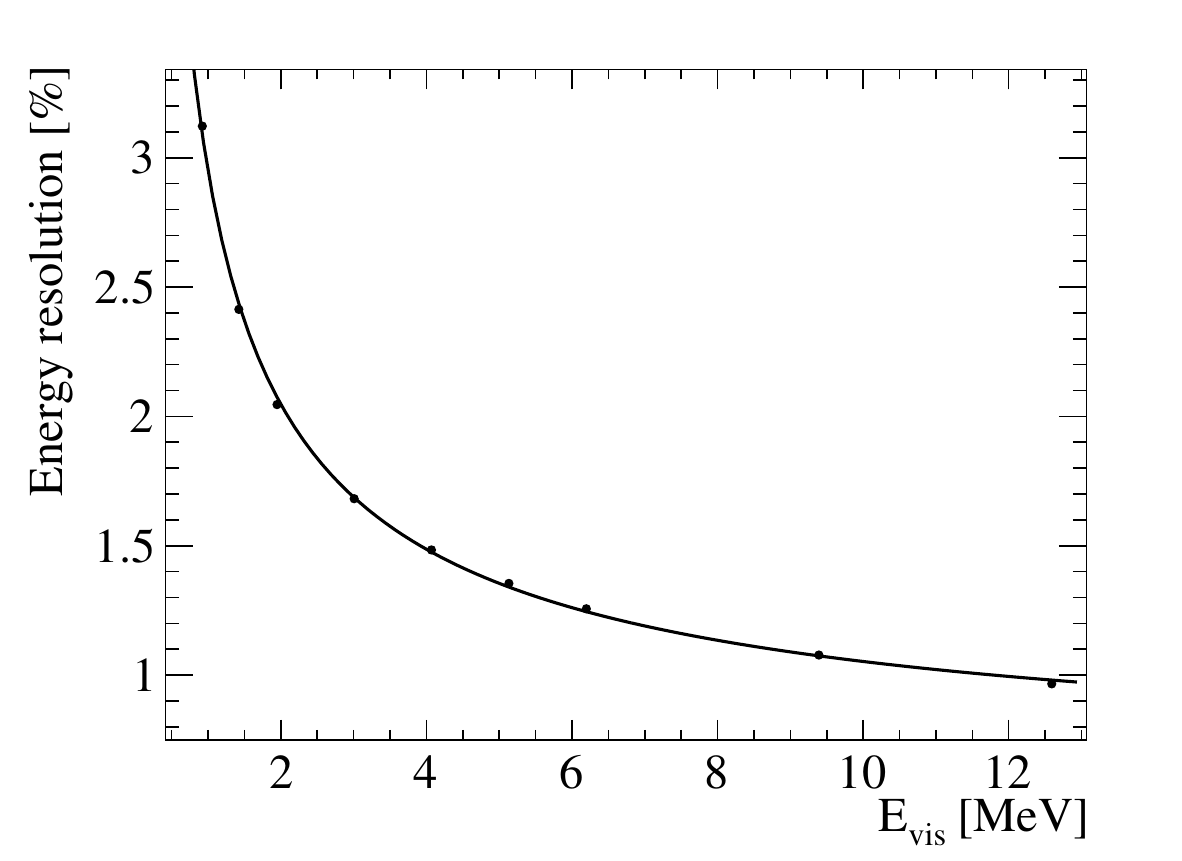}
    \includegraphics[width=0.45\textwidth]{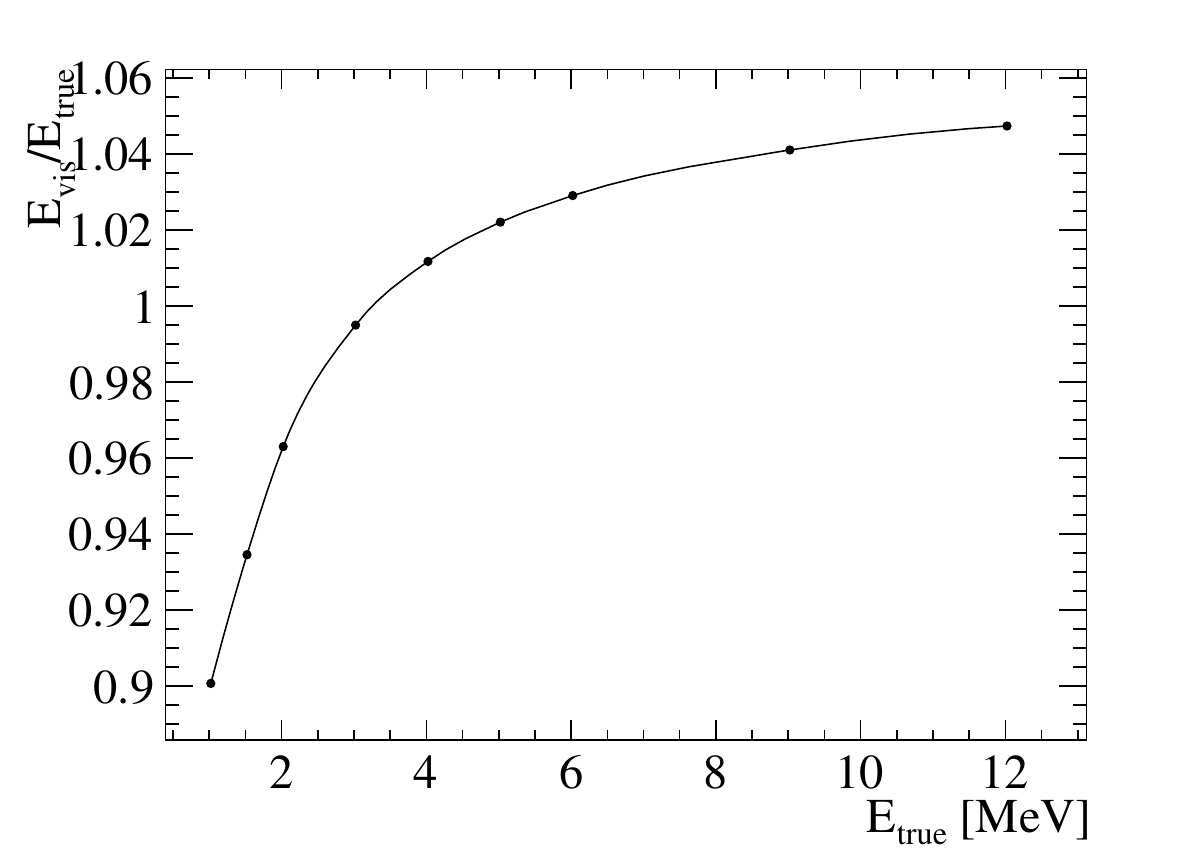}
    \caption{Energy resolution (left) and energy non-linearity (right) for positrons using samples from Tab.~\ref{tab:PhySampleInfo}. A fit with Eq.~\ref{eq:abc_model} was performed for the points in the left plot, while interpolation was used instead in the right plot.}
    \label{fig:rec}
\end{figure}

 In Fig.~\ref{fig:rec}, the left plot shows the energy resolution as a function of the average visible energy $E_\textrm{vis}$. The data points are fitted with a generic parameterization formula as follows: 
\begin{equation}
        \frac{\sigma}{E_\textrm{vis}}=\sqrt{\left(\frac{a}{\sqrt{E_\textrm{vis}}}\right)^2+b^2+\left(\frac{c}{E_\textrm{vis}}\right)^2}. 
\label{eq:abc_model}
\end{equation}
In this equation, $a$ is the statistical term mainly driven by the Poisson statistics of detected PE. The $b$ term is a constant, independent of energy and mostly contributed by the scintillation quenching effect, Cherenkov radiation, and energy non-uniformity. The $c$ term accounts for the PMT dark noise and the positron annihilation $\gamma$s. The best-fit results of $a$, $b$, and $c$ are listed as the Default Case in Tab.~\ref{tab:EresComp54}, and the fitted energy resolution is 2.95\% at 1~MeV. The right plot in Fig.~\ref{fig:rec} shows the energy non-linearity curve.
These results were used for the NMO sensitivity in the JUNO paper ~\cite{JUNO:2024jaw}.

\begin{table}[ht]
\centering
    \caption{Summary of the energy reconstruction results. All the energy units are in MeV. For each set of positrons with different $E_k$, the reconstructed visible energy is fitted with a Gaussian function, where $E_\textrm{vis}$ and $\sigma$ represent the Gaussian mean and sigma, respectively. The energy resolution $E_\textrm{res}$ is defined as $\sigma$/${E_\textrm{vis}}$. In addition, we also report the ratio of the visible energy to the deposited energy.}
\begin{tabular}{cccccccccc}
\toprule
$E_{k}$ [MeV] &  0  & 0.5 &  1  &   2 &  3  &  4  &  5  &   8 & 11 \\
\hline
$E_\textrm{dep}$ [MeV] &  1.022  & 1.522 &  2.022  &   3.022 &  4.022  &  5.022  &  6.022  &  9.022 & 12.022 \\
${E_\textrm{vis}}$ [MeV]  &   0.9205 & 1.422  &  1.947  &   3.007 &  4.069  &  5.133  &  6.197  &  9.392 & 12.59 \\
$E_\textrm{res}$ [\%]  &  3.122  & 2.414 & 2.046  &  1.682 &  1.484  &  1.354  &  1.256  &  1.077 & 0.9661 \\
${E_\textrm{vis}}$/$E_\textrm{dep}$  &   0.901 & 0.934  &  0.963  &  0.995  &  1.012  &  1.022  & 1.029   & 1.041  & 1.047 \\
\bottomrule
\end{tabular}
\label{tab:E_rec}
\end{table}

\begin{figure}
    \centering
    \includegraphics[width=0.45\textwidth]{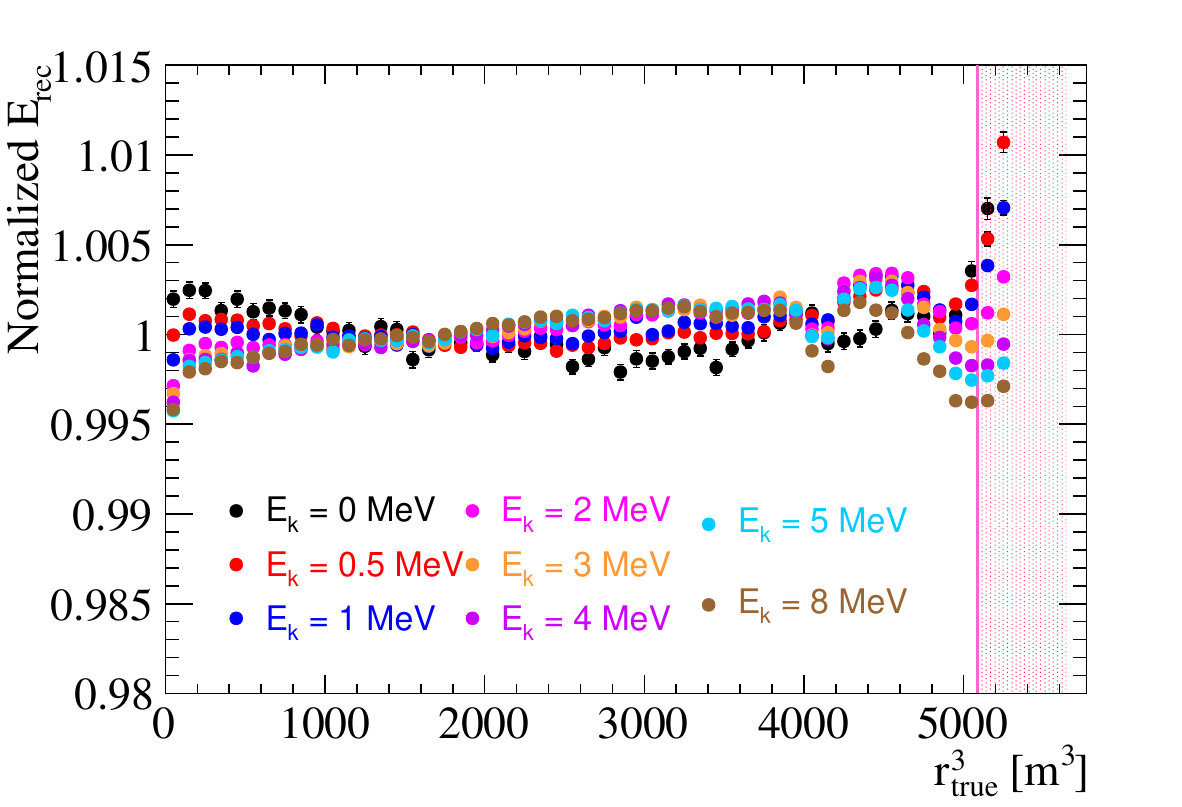}
    \includegraphics[width=0.45\textwidth]{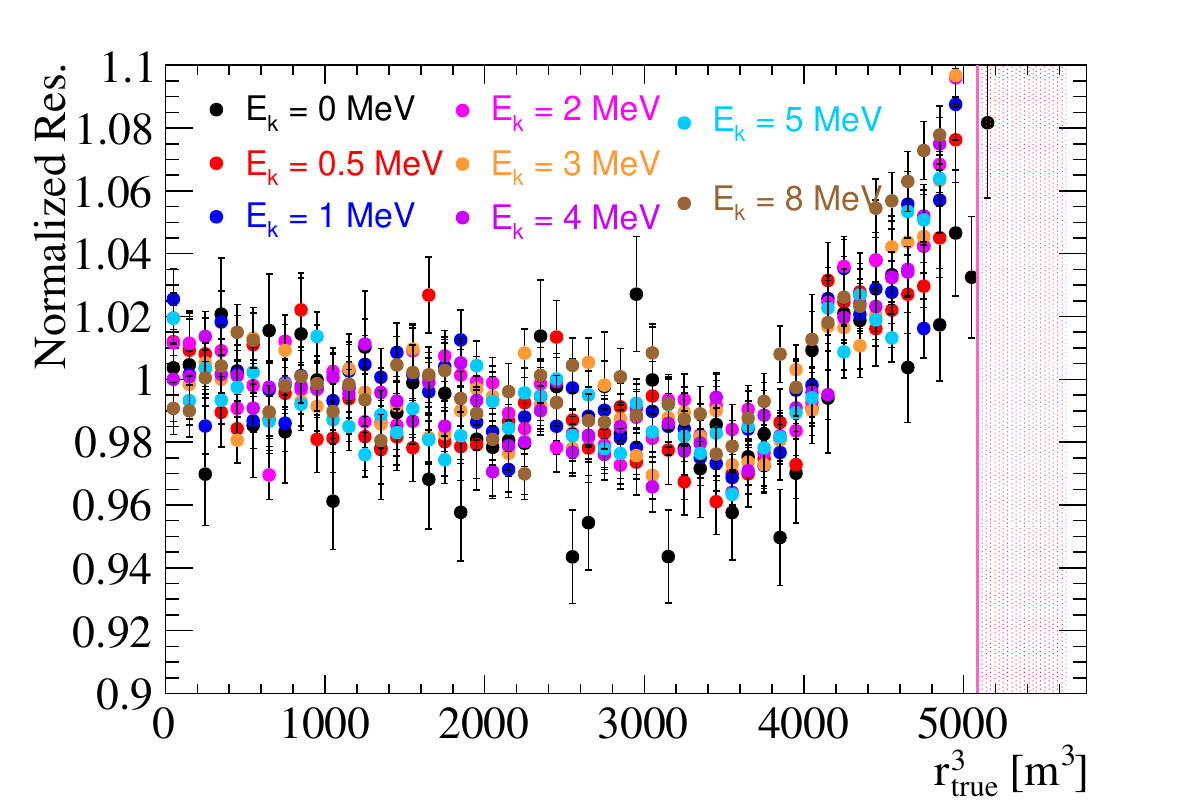}  
    \caption{Energy response non-uniformity checks. Normalized average $E_\textrm{rec}$ (left) and energy resolution (right) with respect to $r^3$ for positron samples with different energies. The normalization is conducted by dividing the average value within the FV. The red vertical line corresponds to the FV cut.}
    \label{fig:nonUniformity1}
\end{figure}

A few additional checks were performed to validate the results. The left plot in Fig.~\ref{fig:nonUniformity1} shows that the energy non-uniformity is within 0.4\% for positrons with different energies. The right plot shows how the energy resolution changes with respect to $r^3$, which is caused mainly by the change in the total number of detected PEs.


\clearpage
\section{Decomposition of energy resolution} 
\label{decomposition}

The contribution of major factors to the energy resolution budget is estimated at two different phases: detector simulation phase and calibration and reconstruction process. In the detector simulation phase, positron samples are generated at the CD center. However, in the reconstruction process, the positron vertexes are generated with a uniform distribution within the LS volume.

\subsection{Decomposition of energy resolution in the detector simulation}

\begin{figure}
    \centering
    \includegraphics[scale=0.5]{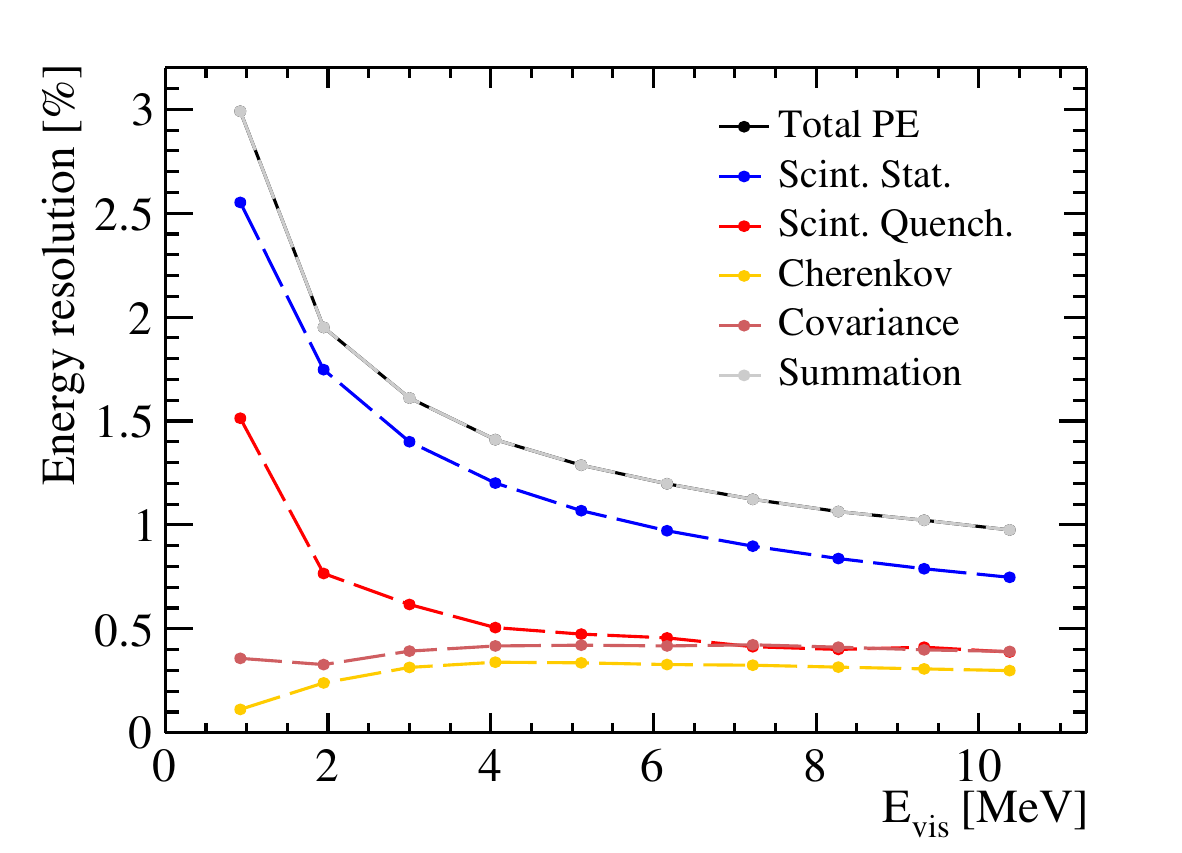}
    \caption{Energy resolution decomposition. The black curve represents the energy resolution obtained with total PE. The blue, red, dark red, and yellow curves represent the contributions from scintillation Poisson statistics, quenching effect, Cherenkov radiation and the covariance between scintillation, and Cherenkov processes, respectively. The gray curve, which superimposes with the black curve, is the root squared sum of the four components.}
    \label{fig:resolution_decomposition_detsim}
\end{figure}

The positron energy resolution at the center of the CD is obtained from  detector simulation. According to the recorded truth information, the contributions from scintillation light, Cherenkov light, and their covariance to the energy resolution are extracted, and major effects in the energy resolution budget are shown in Fig.~\ref{fig:resolution_decomposition_detsim}. It has previously been discussed that the quenching effect causes the number of scintillation PEs to deviate from a Poisson distribution. Therefore, the contribution from scintillation PEs can be further decomposed into two parts: the Poisson fluctuation (blue curve) and quenching effect (red curve). The contribution from Poisson fluctuation is evaluated using the square root of the number of scintillation PEs, which is given by $\sigma_\textrm{stat} = \sqrt{N_\textrm{ScintPE}}$. This component represents the statistical fluctuations in the scintillation process. The contribution from the quenching effect is then obtained by subtracting the Poisson standard deviation from the standard deviation of the scintillation PE number distribution. Mathematically, this is expressed as $\sigma_\textrm{quench} = \sqrt{\sigma_\textrm{ScintPE}^2 - \sigma_\textrm{stat}^2}$, where $\sigma_\textrm{ScintPE}$ is the standard deviation of the scintillation PE number distribution. In Fig.~\ref{fig:resolution_decomposition_detsim}, the yellow curve represents the contribution from Cherenkov radiation, and the dark red curve represents the correlation between the scintillation and Cherenkov processes. As expected, the gray curve, which is the square root of the quadratic sum of the four components, accurately reproduces the total energy resolution (black curve) obtained directly from the standard deviation of the total PE spectrum.

It is worth noting that the standard deviation of the total PE is significantly larger than the Poisson standard deviation, indicating the presence of significant systematic effects in the energy resolution beyond statistical fluctuations. This highlights the importance of considering these systematic effects in the analysis of the energy resolution in the detector simulation.

\subsection{Decomposition of energy resolution in the reconstruction}
There are a few key factors that could affect the energy resolution in the event reconstruction. Due to energy non-uniformity, the vertex uncertainty will propagate to the energy resolution. A more precise vertex could improve the energy resolution. 
Another important contributor comes from the PMT dark noise, which contaminates the signal photons and worsens the energy resolution. 
Moreover, the inputs to the event reconstruction, namely, the charge and time of PMTs, are obtained from the PMT waveform reconstruction, charge non-linearity and waveform reconstruction uncertainty, will propagate to the energy resolution as well. 
Finally, the PMT charge could only provide a rough estimation of the number of detected PEs. The intrinsic PMT charge resolution, which includes contributions from both Gaussian and exponential components (Sec.~\ref{sec:lpmt_response}), will lead to charge smearing and potentially worsen the energy resolution for the charge-based energy reconstruction.
The impact of these factors on the energy resolution is decomposed in this section.
Different cases were considered, in which the factors above were removed sequentially, as follows:
\begin{itemize}
    \item Default Case: vertex and energy are simultaneously reconstructed.
    \item Case A: true vertex is used in the energy reconstruction.
    \item Case B: PMT dark noise is removed in the samples and true vertex is used.
    \item Case C: on top of Case B, the waveform reconstruction is replaced by a toy simulation to provide the PMT charge and time in the reconstruction. 
    \item Case D: in addition to the changes in Case C, SPE charge resolution and other electronics effects are also removed.
\end{itemize}
In the default case, the positron vertex is simultaneously reconstructed with its energy using the method described in Sec.~\ref{sec:rec_algo}. 
The vertex resolution is approximately 10~cm at 1~MeV and decreases at higher energies. Meanwhile, the vertex bias is less than 2~cm within the FV. 
In case A, everything is the same as in the default case, except that the MC truth vertex is used in the energy reconstruction, which is equivalent to a vertex resolution of 0~mm. 
Besides using the true vertex, the PMT dark noise is removed in case B. 
Given that there is no more dark noise, the 420~ns signal window is discarded in both the nPE map and energy reconstruction.
In case C, the waveform reconstruction is replaced by a toy simulation that smears the charge of each PE, providing both charge and time information.
This smeared PMT charge from the toy simulation is used in both the nPE map and the energy reconstruction. In comparison, the charge reconstructed from PMT waveforms is used in previous cases. 
By using these toy electronic simulation samples, we are able to remove the impact from the waveform reconstruction as well as other electronics effects, except the PMT SPE charge smearing. The true vertex is used for case C.
For case D, the PMT SPE charge smearing is removed on top of case C by directly using the detector simulation samples for the reconstruction. Consequently, the new observable is the PE number, instead of the reconstructed charge. This corresponds to the most ideal scenario and leads to the best energy resolution. 

\begin{figure}
    \centering
    \includegraphics[width=0.49\textwidth]{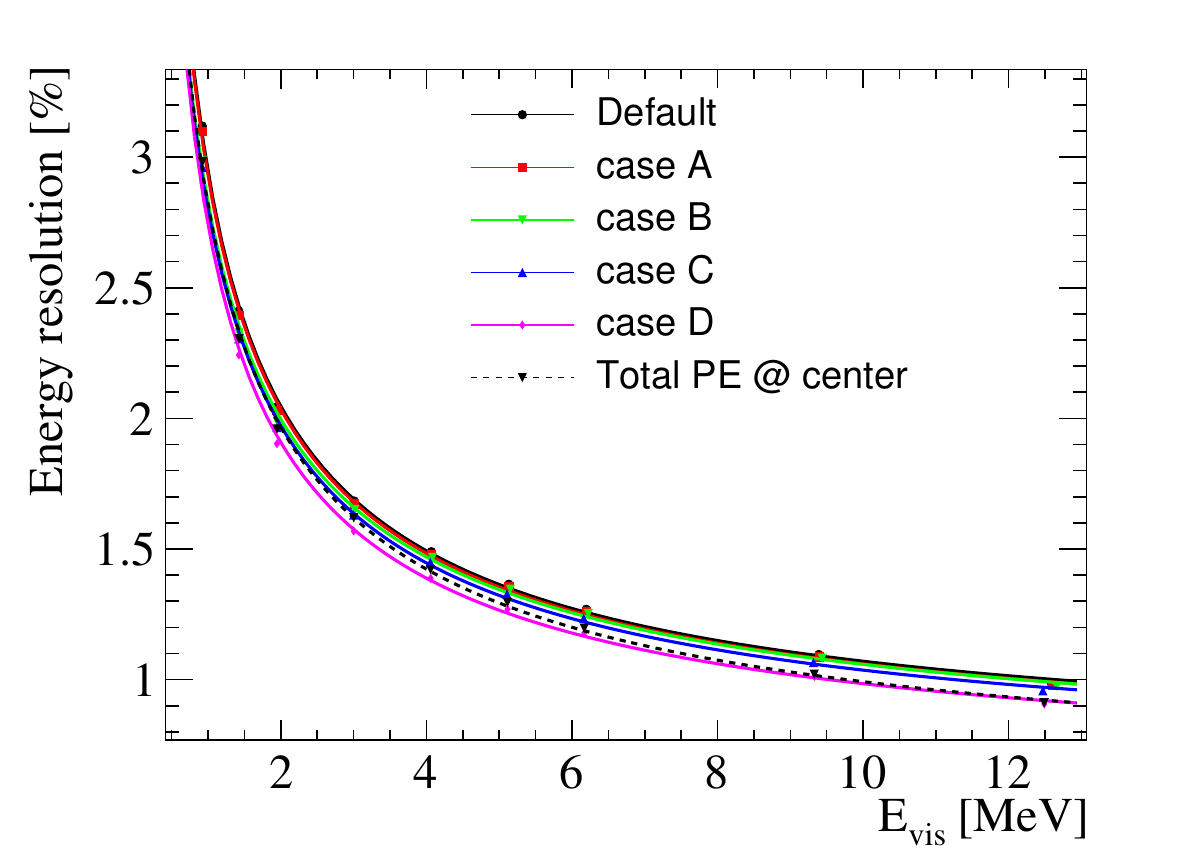}
    \includegraphics[width=0.49\textwidth]{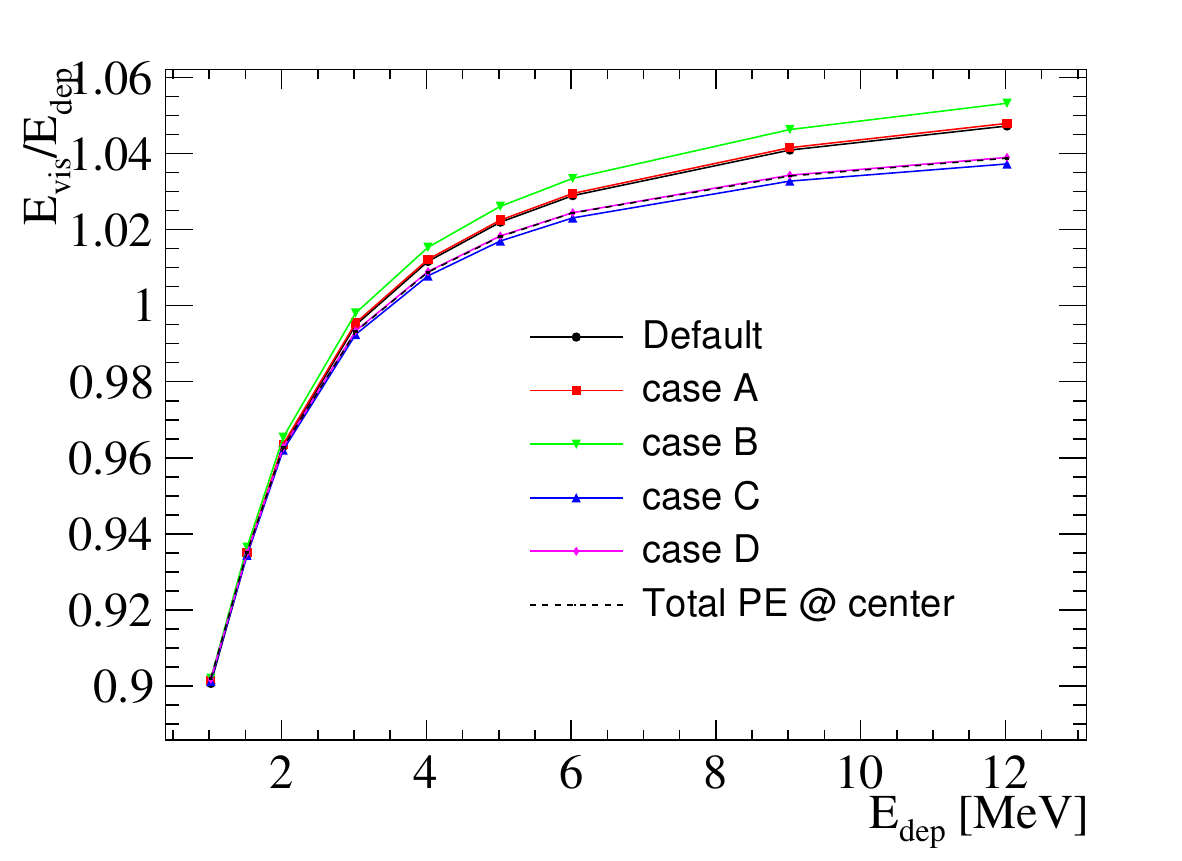}
    \caption{Comparison of the energy resolution (left) and energy non-linearity (right) among the different cases. The energy resolution for each case is better than that of the previous one, indicating that the energy resolution improves by moving from the real to ideal situation. The black curve from Fig.~\ref{fig:resolution_decomposition_detsim} is represented by the dashed curve here, and its corresponding energy resolution is close to that of case D.}
    \label{fig:EresNL_compare54}
\end{figure}


Figure~\ref{fig:EresNL_compare54} shows a comparison of the energy resolution and energy non-linearity among all cases. We can see that the energy resolution improves monotonically among the cases, indicating that the energy resolution keeps improving by removing each contributing factor.
The fitting results of the energy resolution for all cases are summarized in Tab.~\ref{tab:EresComp54}. 
The relative improvement of the energy resolution at 1~MeV for each case with respect to the previous case is also presented.
We can see that at 1~MeV, the PMT dark noise and SPE charge smearing are the two dominant contributing factors to the energy resolution, with relative improvements of approximately 3.45\% and 2.1\%, respectively.
Removing the charge uncertainty from the PMT waveform reconstruction together with charge non-linearity leads to a 0.92\% relative improvement.
The impact of vertex resolution is relatively small, and the relative improvement is approximately ~0.78\% using the true vertex.
The black curve from Fig.~\ref{fig:resolution_decomposition_detsim} is also shown as the dashed curve here, and its corresponding energy resolution is close to that of case D.

\begin{table}[!ht]
\centering
\caption{Comparison of the energy resolutions among all the cases. Here a, b and c correspond to the three parameters from Eq.~\ref{eq:abc_model}. The relative improvement of the energy resolution at 1~MeV with respect to the previous case is also shown in the last column.}
    \begin{tabular}{lccccc}
\toprule
       Case     & \textbf{a} [\%] & \textbf{b} [\%] & \textbf{c} [\%] & $E_\textrm{res}$@1~MeV [\%]  & Sequential improvement\\
\midrule
      Default & 2.614 & 0.640 & 1.205 & 2.948 & - \\
            A (- vertex uncert.)       & 2.581 & 0.667 & 1.206 & 2.925 &  0.78\% \\
            B (- dark noise)       & 2.571 & 0.671 & 0.956 & 2.824 & 3.45\% \\
            C (- waveform reco)       & 2.542 & 0.647 & 0.973 & 2.798 &  0.92\% \\
            D (- SPE charge smear)       & 2.445 & 0.600 & 1.079 & 2.739 & 2.1\% \\
\bottomrule

\end{tabular}
\label{tab:EresComp54}
\end{table}

\begin{figure}
    \centering
    \includegraphics[width=0.8\textwidth]{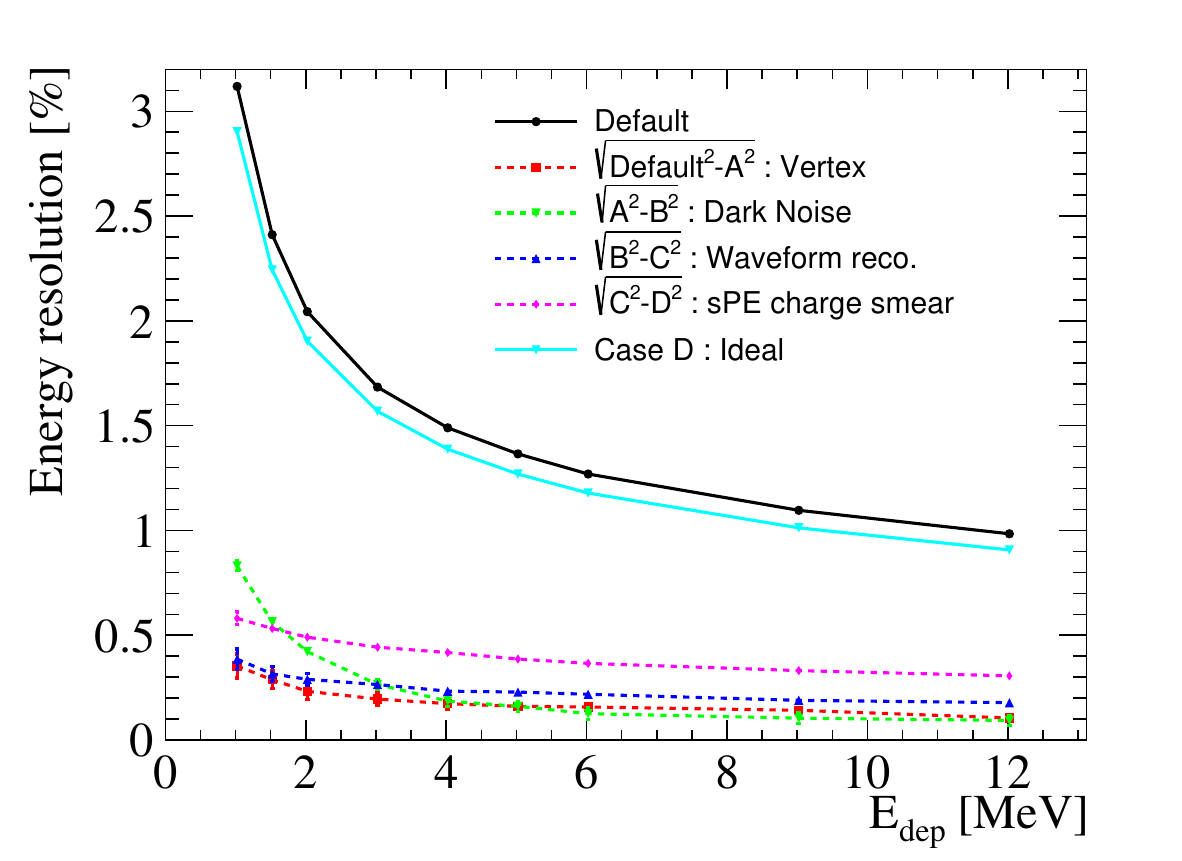}    
    \caption{Decomposition of the energy resolution. The black solid curve represents the default case. It is decomposed into five parts: the four dotted curves correspond to the contributions of vertex resolution (red), dark noise (green), waveform reconstruction (blue), and SPE charge smear (purple), and the light blue solid curve represents case D (ideal case). 
    Taking 1.022~MeV as an example, the values for all curves are 3.12\%, 0.35\%, 0.83\%, 0.39\%, 0.58\% and 2.90\%, and the decomposition can be written as $3.12^2 = 0.35^2 + 0.83^2 + 0.39^2 + 0.58^2 + 2.90^2$.}
    \label{fig:Eres_decompose}
\end{figure}

The above comparison shows the average impact of each factor on the energy resolution. 
In addition, it also enables us to decompose the energy resolution at each discrete energy point.
With the energy resolution curves of all cases, the default energy resolution can be decomposed into five components, as shown in Fig.~\ref{fig:Eres_decompose}. 
The black solid curve corresponds to the default case, which is used for the NMO analysis~\cite{JUNO:2024jaw}. The teal solid curve corresponds to case
D and represents the ideal energy resolution we could obtain.
The four dashed curves show the contribution of each factor to the energy resolution, and the values at each energy point are calculated as $\sqrt{(E^{X}_\textrm{res})^2 - (E^{Y}_\textrm{res})^2}$, 
where Y goes from case A to D and X corresponds to the preceding case. 
Two features stand out immediately. The contribution from dark noise decreases as the energy increases, as expected. 
Moreover, in the energy range below 1.5~MeV, dark noise is the major contributor. 
For the energy range above 1.5~MeV, particularly the most sensitive region of (1.5, 3)~MeV for the NMO analysis, the SPE charge smearing is the dominant factor. 
Preliminary studies indicate that its impact could be partially mitigated by PE counting, especially for PMTs with 1 or 2 PEs. 

\section{Summary} \label{sec:summary}
The energy resolution is a crucial parameter in determining the sensitivity of the JUNO experiment to NMO. However, in addition to the statistical fluctuations in the detected number of PEs, several other effects impact the energy resolution in the detection of IBD signals. This paper presented a comprehensive study of the energy resolution in the JUNO experiment, incorporating the latest knowledge and updates in the detector construction stage. This includes a better understanding of the detector structures, more precise measurements of the optical properties of the LS, comprehensive evaluations of the characteristics of both LPMTs and SPMTs, improved modeling of the spectral and angular dependencies of the LPMTs' PDE, and better constraints on the absolute scintillation and Cherenkov light yield based on data from the Daya Bay experiment. All of these updates have been implemented in the JUNO detector simulation, which also includes detailed modeling of the PMT responses and readout electronics in the electronics simulation.

By using data samples generated with the full detector and electronics simulation, a full-chain data processing of calibration and reconstruction was performed to evaluate factors that can smear the energy resolution. These factors include residual energy non-uniformity after reconstruction, accuracy of reconstructed energies from reconstruction algorithms, impacts of dark count rate, and SPE charge resolution. After considering these factors, it was found that an overall energy resolution of 2.95\% at 1 MeV can be achieved for positrons from IBD signals, and the obtained energy resolution curve has been applied in the recent NMO analysis of JUNO. Furthermore, the contribution of major effects in the energy resolution budget was estimated. This study serves as a reference for interpreting future measurements of energy resolution in JUNO data collection and provides a guideline for understanding the energy resolution of LS-based detectors. 

After the data collection phase of the JUNO experiment begins, further updates and improvements are expected in the understanding of the energy resolution. The analysis of the collected data will provide valuable insights and allow for refinements in the modeling and calibration of the detector. These updates and improvements will contribute to the ongoing efforts to optimize the performance of the JUNO detector and enhance its sensitivity to NMO.

\section*{Acknowledgement}

We acknowledge the Daya Bay collaboration for providing the Monte Carlo simulation software.
We are grateful for the ongoing cooperation from the China General Nuclear Power Group.
This work was supported by
the Chinese Academy of Sciences,
the National Key R\&D Program of China,
the CAS Center for Excellence in Particle Physics,
Wuyi University,
and the Tsung-Dao Lee Institute of Shanghai Jiao Tong University in China,
the Institut National de Physique Nucl\'eaire et de Physique de Particules (IN2P3) in France,
the Istituto Nazionale di Fisica Nucleare (INFN) in Italy,
the Italian-Chinese collaborative research program MAECI-NSFC,
the Fond de la Recherche Scientifique (F.R.S-FNRS) and FWO under the ``Excellence of Science – EOS” in Belgium,
the Conselho Nacional de Desenvolvimento Cient\'ifico e Tecnol\`ogico in Brazil,
the Agencia Nacional de Investigacion y Desarrollo and ANID Millennium Science Initiative Program — ICN2019\_044 in Chile,
the Charles University Research Centre and the Ministry of Education, Youth, and Sports in Czech Republic,
the Deutsche Forschungsgemeinschaft (DFG), the Helmholtz Association, and the Cluster of Excellence PRISMA+ in Germany,
the Joint Institute of Nuclear Research (JINR) and Lomonosov Moscow State University in Russia,
the joint Russian Science Foundation (RSF) and National Natural Science Foundation of China (NSFC) research program,
the MOST and MOE in Taiwan,
the Chulalongkorn University and Suranaree University of Technology in Thailand,
and the University of California at Irvine and the National Science Foundation in USA.

\bibliographystyle{h-physrev5}
\bibliography{references}

\end{document}